\pdfoutput=1
\newcommand*{\ATLASLATEXPATH}{}
\documentclass[PAPER, texlive=2016, UKenglish, cernpreprint]{\ATLASLATEXPATH atlasdoc}
 
\usepackage[subfigure]{\ATLASLATEXPATH atlaspackage}
 
\usepackage{\ATLASLATEXPATH atlasbiblatex}
 
\usepackage{\ATLASLATEXPATH atlasphysics}
 
\newcommand*{\numRF}[2]{\num[round-mode=figures,round-precision=#2]{#1}}

\graphicspath{{logos/}{figures/}}
 
\usepackage{ANA-STDM-2017-03-PAPER-defs}
 
\usepackage[utf8]{inputenc}
\usepackage{multirow}

\AtlasTitle{Measurement of $ZZ$ production in the \llnn~final state
with the ATLAS detector in $pp$ collisions at $\sqrt{s} = $ 13 \TeV}

\AtlasAbstract{
This paper presents a measurement of $ZZ$ production with the ATLAS detector at the Large Hadron Collider.
The measurement is carried out in the final state with two charged leptons and two neutrinos,
using data collected during 2015 and 2016 in $pp$ collisions at $\sqrt{s} = $ 13 \TeV, corresponding
to an integrated luminosity of 36.1 fb$^{-1}$.
The integrated cross-sections in the total and fiducial phase spaces are measured with an uncertainty of 7\%
and compared with Standard Model predictions, and differential measurements in the fiducial phase space are reported.
No significant deviations from the Standard Model predictions are observed,
and stringent constraints are placed on anomalous couplings corresponding to neutral triple gauge-boson interactions.
}
 
\author{The ATLAS Collaboration}
 
\AtlasRefCode{STDM-2017-03}
 
\PreprintIdNumber{CERN-EP-2019-066}
 
\AtlasDate{\today}

\AtlasJournalRef{JHEP 10 (2019) 127}
\AtlasDOI{10.1007/JHEP10(2019)127}

\AtlasCoverSupportingNote{Support Note}{https://cds.cern.ch/record/2239641}

\AtlasCoverCommentsDeadline{April 9$^\text{th}$ 2019}
 
\AtlasCoverAnalysisTeam{
James Cowley, Cong Geng, Wen Guo, Liang Han, Suen Hou,
Bo Liu, Yusheng Wu, Guangyi Zhang, Bing Zhou}
 
\AtlasCoverEdBoardMember{Donatella Cavalli, Daniel Froidevaux (chair), Vasiliki Kouskoura}

\AtlasCoverEgroupEditors{atlas-stdm-2017-03-editors@cern.ch}
 
\AtlasCoverEgroupEdBoard{atlas-stdm-2017-03-editorial-board@cern.ch}
 
\hypersetup{pdftitle={ATLAS document},pdfauthor={The ATLAS Collaboration}}
 
\begin{document}
 
\maketitle
 
\tableofcontents

\section{Introduction}
\label{sec:intro}
 

In the Standard Model (SM), the production of gauge boson pairs has a profound connection with
the non-Abelian nature of the electroweak (EW) theory and with the spontaneous breaking of the EW gauge symmetry.
In addition, a broad range of new phenomena beyond the SM (BSM) are predicted to reveal themselves through diboson production.
The study of diboson production probes
a cornerstone of the EW theory and possible BSM physics scenarios, and it constitutes a salient component of
the physics programme at the Large Hadron Collider (LHC). Among all the diboson processes,
the production of two on-shell $Z$ bosons has the smallest cross-section,
but is nevertheless quite competitive for measurements and searches,
because of its generally good signal-to-background ratio for the fully leptonic decay channels.
For instance, the $ZZ$ process is a leading channel to search for anomalous neutral
triple-gauge-boson couplings (aTGCs)~\cite{Baur:2000ae} and to study the off-shell production of
the Higgs boson~\cite{Goncalves:2017iub,Offshell_ATLAS_13TeV}.

\begin{figure}[t]
\centering
\subfigure[]{\includegraphics[width=0.33\figwidth]{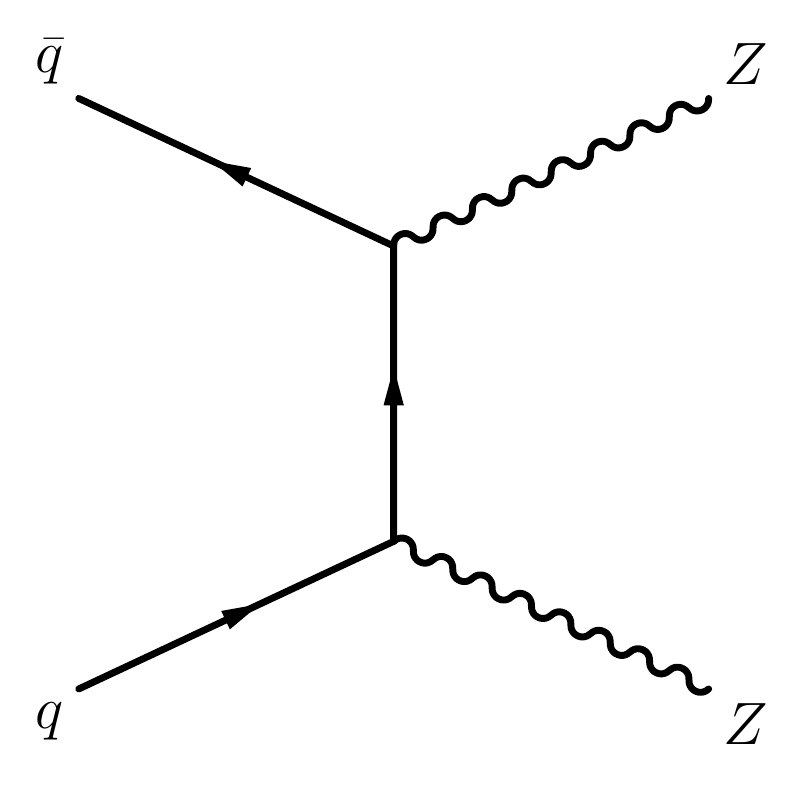}}
\subfigure[]{\includegraphics[width=0.33\figwidth]{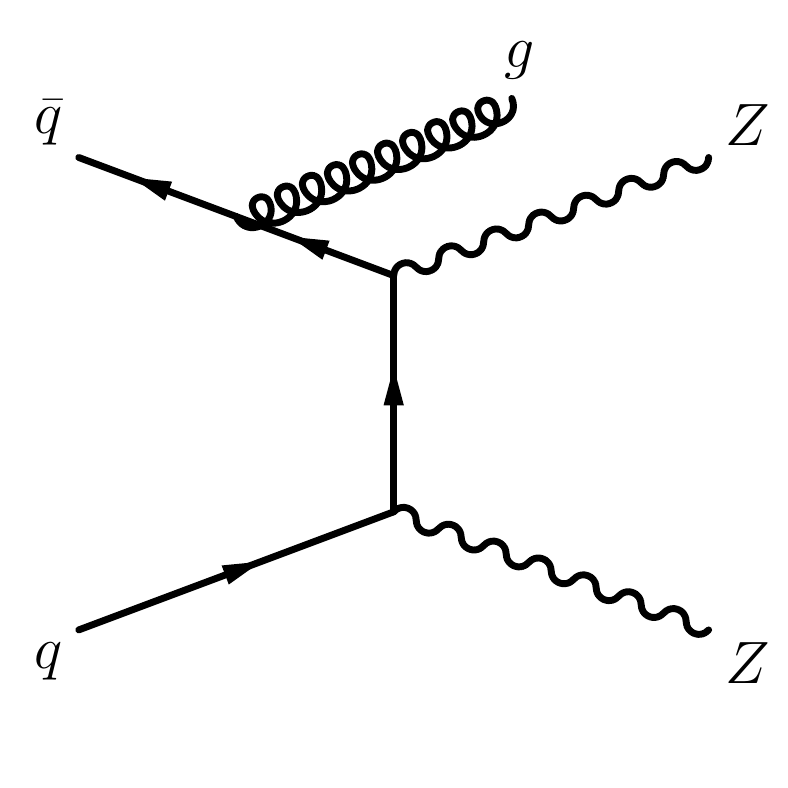}}
\subfigure[]{\includegraphics[width=0.33\figwidth]{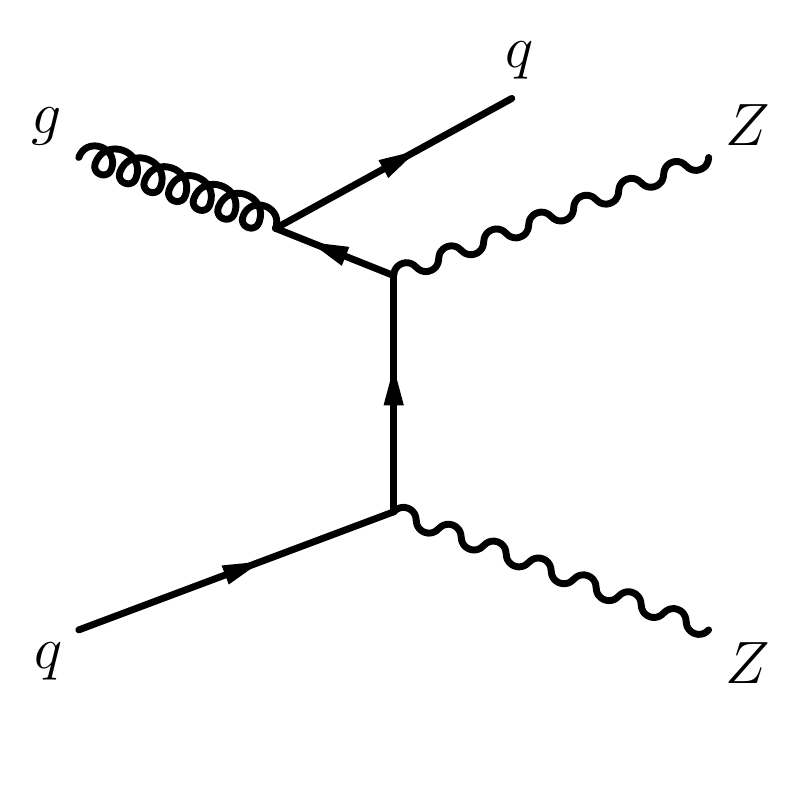}} \\
\subfigure[]{\includegraphics[width=0.33\figwidth]{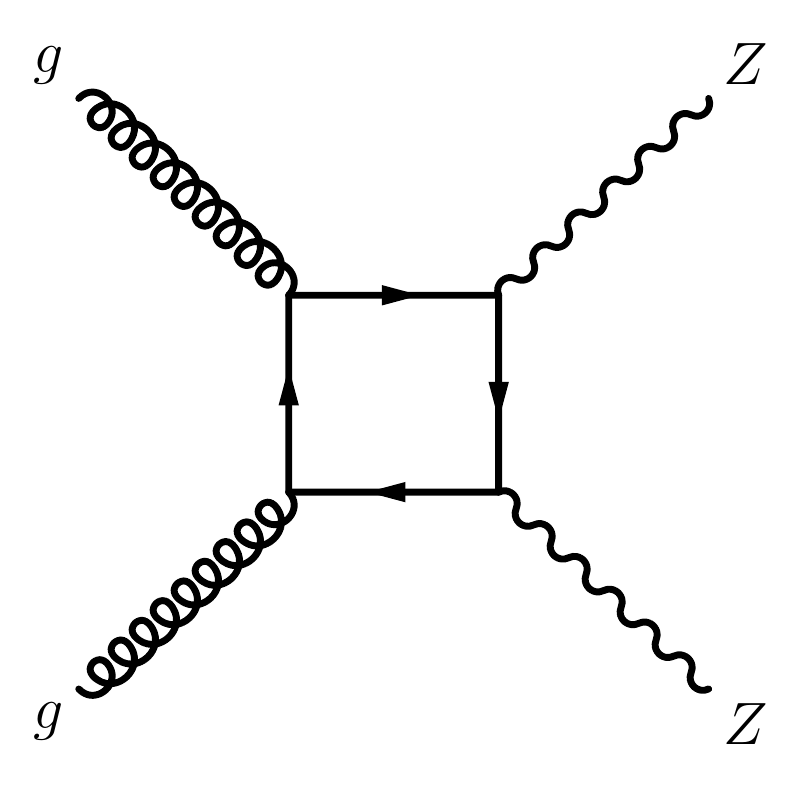}}
\subfigure[]{\includegraphics[width=0.33\figwidth]{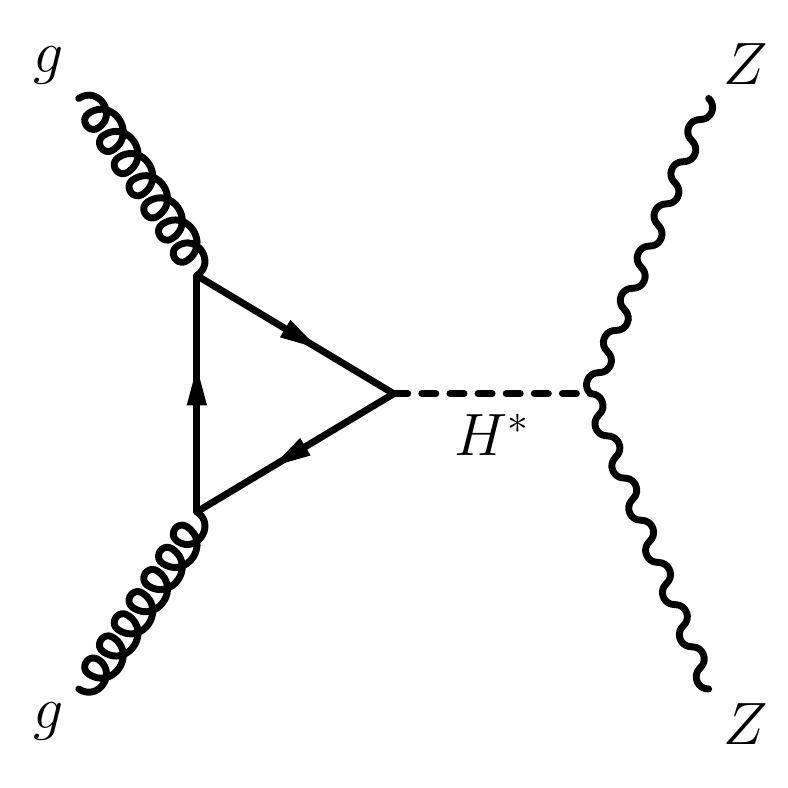}}
\subfigure[]{\includegraphics[width=0.33\figwidth]{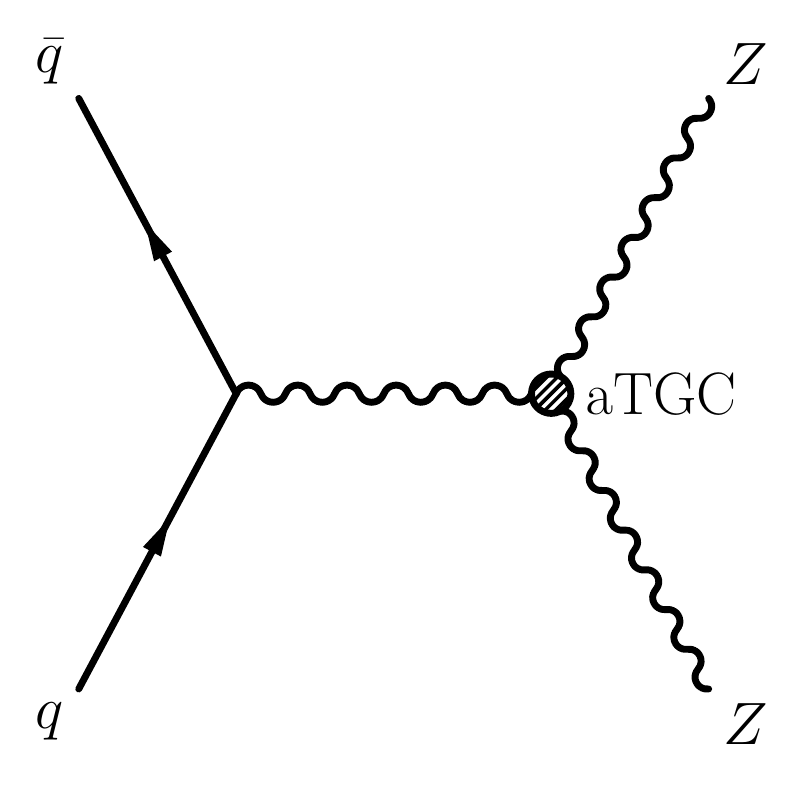}}
\caption{Representative Feynman diagrams for $ZZ$ production at the LHC:
(a) lowest-order $t$-channel $qqZZ$ production;
(b) production of $ZZ$ plus one parton through the $q\bar{q}$ initial state;
(c) production of $ZZ$ plus one parton through the $qg$ initial state;
(d) $ggZZ$ production with a fermion loop;
(e) $ggZZ$ production involving an exchange of a virtual Higgs boson;
(f) $s$-channel production with aTGCs.
}
\label{fig:diagram}
\end{figure}

\Fig{\ref{fig:diagram}} shows representative Feynman diagrams for $ZZ$ production at the LHC.
The dominant process is $t$-channel production with a quark and anti-quark initial state,
hereafter denoted by the $qqZZ$ process. Higher-order QCD corrections to the $qqZZ$ process are found to
be sizeable~\cite{Cascioli:2014yka}, and two tree-level diagrams concerning production of two $Z$ bosons and
one outgoing parton are shown. The gluon fusion process ($ggZZ$) includes two sub-processes,
one with a fermion loop and the other involving a virtual Higgs boson.
Although the $ggZZ$ process only appears at $O(\alpha_{\mathrm{S}}^{2})$, it nevertheless has a non-negligible
contribution of $O(10\%)$ to the total $ZZ$ production rate due to the large gluon flux at the LHC.
The $s$-channel production is forbidden at the lowest order; however, the neutral TGCs can still acquire small values of $O(10^{-4})$ in the SM,
due to the correction with a fermion loop~\cite{Gounaris:2000tb}.
The observation of aTGCs with larger values would hint at the existence of new physics.

Measurements of $ZZ$ production at the LHC have been carried out in two decay final states, one with four charged leptons ($4\ell$)
and the other with two charged leptons and two neutrinos (\llnn).
Using LHC Run-1 and Run-2 data, multiple results~\cite{STDM-2012-02,STDM-2014-16,STDM-2016-15,CMS-SMP-12-007,CMS-SMP-12-016,CMS-SMP-13-005,CMS-SMP-16-017} have been reported by the ATLAS and CMS experiments.
The most precise results to date have been obtained from the $4\ell$ channel using 13 \TeV{} data~\cite{STDM-2016-15,CMS-SMP-16-017},
where the integrated production cross-section has been measured to a precision of 5\% and the upper bound on neutral aTGC parameters
has been reduced to $10^{-3}$.
The improved experimental precision has stimulated theoretical calculations with a greater accuracy,
and the next-to-next-to-leading-order (NNLO) QCD~\cite{Cascioli:2014yka,Grazzini:2015hta,Grazzini:2017mhc,Heinrich:2017bvg,Kallweit:2018nyv} and
next-to-leading-order (NLO) EW~\cite{NLOEWcorr,Kallweit:2017khh} predictions have become available for the $qqZZ$ process.

This paper presents a measurement of $ZZ$ production using 36.1 \ifb~of data collected with the ATLAS detector in $pp$ collisions at $\sqrt{s} = $ 13 \TeV. This analysis is performed in the \llnn~($\ell = e$ or $\mu$) final state, which has a larger branching fraction but suffers from
higher background contamination in comparison with the $4\ell$ channel.
To ensure a good signal-to-background ratio, the experimental selection requires
one $Z$ boson boosted against the other in the transverse plane, which results in a pair of high-\pt{}
isolated leptons and significant missing transverse momentum (\met).
The \llnn{} channel thus offers higher data statistics than the $4\ell$ channel for events with high-\pt $Z$
bosons, and offers competitive precision for integrated and differential measurements, as well as good sensitivity to aTGCs.

The dominant background arises from $WZ$ production where the $Z$ boson decays into a pair of charged leptons.
About 60\% of the $WZ$ events which contribute to the \llnn~final state have the $W$ boson decaying leptonically ($W\rightarrow\ell\nu$ or $W\rightarrow\tau\nu\rightarrow\ell+3\nu$, $\ell = e$ or $\mu$), where the final-state lepton escapes detection. The remaining 40\% $WZ$ contribution is related to the $W\rightarrow\tau\nu$ decay with subsequent hadronic decays of the $\tau$-lepton.
Another important background comes from the processes that genuinely produce the \llnn~final state but
contain a lepton pair not originating directly from a $Z$-boson decay. This background, referred to as the \nonresll~background,
includes $WW$, top-quark (\ttbar~and $Wt$), and \Ztt{} production.
The production of a $Z$ boson in association with jet(s) (\zjet) also constitutes a potentially large background source.
The \zjet{} events with large ``fake'' \met{} arise from heavy-flavour hadron decays in the accompanying jet(s),
from jet mismeasurements in certain regions of the detector, and from the measurement resolution itself, owing to
the additional $pp$ collisions in the same or neighbouring proton bunch crossings (pile-up).
The \ZZllll{} process yields a small contribution when one lepton pair misses detection, and it is considered as a background in this measurement. Finally, minor background contributions are expected from
three-boson production ($VVV$ with $V = W$ or $Z$) and production of \ttbar~accompanied by one or two vector bosons ($t\bar{t}V$).

The integrated cross-section of $ZZ$ production is measured in a fiducial phase space and then extrapolated to a total phase space. The determination of the fiducial (\xsfid) and total (\xstot) cross-sections is obtained as shown in \Eqn{(\ref{eq:xs})}:
 
\begin{equation}
\xsfidEQ = \frac{\nZZobsEQ}{\mathcal{L} \times \cfactorEQ}, ~~~~~
\xstotEQ = \frac{\nZZobsEQ}{\mathcal{L} \times \cfactorEQ \times \afactorEQ \times B} ~,
\label{eq:xs}
\end{equation}
 
where \cfactor{} stands for an overall efficiency correction factor, \afactor{} is the fiducial acceptance,
and $B$ is the branching fraction of the \ZZllnn{} ($\ell=e,\mu$) decay. The signal yield \nZZobs{} is determined through a fit to
the observed \met{} spectrum, which leads to improved sensitivity compared with a simple event-counting method.
The \afactor{} (\cfactor) factor is calculated as $\nZZexpfidEQ/\nZZexptotEQ$ ($\nZZexpdetEQ/\nZZexpfidEQ$),
where \nZZexpdet, \nZZexpfid, and \nZZexptot{} correspond to the expected signal yields for the \ZZllnn{} final state after the detector-level selection,
in the fiducial region, and in the total phase space, respectively.
The definitions of the total and fiducial phase spaces are elaborated in \Sect{\ref{sec:ps}}.
The simulated events arising from the $ZZ\to \tau\tau\nu\nu$ decays with the subsequent $\tau\to\ell\nu\nu$ decays of both $\tau$-leptons are considered
as signal events at detector level but excluded in the fiducial measurements.
Throughout this paper, ``$Z\to\ell\ell$'' denotes the decays of a $Z$ boson or a virtual photon into a charged-lepton pair.
 
Furthermore, differential cross-sections are reported in the fiducial region for eight kinematic variables,
which are sensitive to effects from higher-order corrections and possible BSM physics.
These variables include the transverse momentum of the leading lepton (\ptleadl), the leading jet (\ptleadj),
the dilepton system (\ptll), and the $ZZ$ system (\ptZZ), the transverse mass of the $ZZ$ system
(\mtZZ),\footnote{$m_{\mathrm{T}}^{ZZ} = \sqrt{ \left[ {\sqrt{\mZ^{2} + \left(\pt^{\ell\ell}\right)^2} + \sqrt{\mZ^{2} + \left( \met \right)^2}}\: \right]^2 - \left\lvert \vecptll  + \vec{E}_{\mathrm{T}}^{\mathrm{miss}} \right\rvert^2 }$}
the absolute rapidity of the dilepton system (\yllabs), the azimuthal angle difference between the two leptons (\dphill),
and the number of jets (\njets).
Since no significant deviations from the SM are observed, upper limits are placed on the aTGC parameters~\cite{Baur:2000ae}, which
typically manifest themselves as a signal excess growing rapidly
as the partonic centre-of-mass energy (\shat{}) increases. In this analysis, aTGCs are searched for using the \ptll{} spectrum in the fiducial region, motivated by the fact that \ptll{} is correlated with \shat{} and has a good experimental resolution.
 

\section{ATLAS detector}
\label{sec:det}
 
 
The ATLAS detector~\cite{Aad:2008zzm,ATLAS-TDR-19,Abbott:2018ikt} is a large multi-purpose detector with a forward--backward symmetric cylindrical geometry and nearly 4$\pi$ coverage in solid angle.\footnote{ATLAS uses a right-handed coordinate system with its origin at the nominal interaction point (IP) in the centre of the detector and the $z$-axis along the beam pipe. The $x$-axis points from the IP to the centre of the LHC ring, and the $y$-axis points upward. Cylindrical coordinates $(r,\phi)$ are used in the transverse plane, $\phi$ being the azimuthal angle around the $z$-axis. The pseudorapidity is defined in terms of the polar angle $\theta$ as $\eta=-\ln\tan(\theta/2)$.}
It consists of an inner tracking detector surrounded by a thin superconducting solenoid, electromagnetic and hadronic calorimeters, and a muon spectrometer incorporating three large superconducting toroidal magnets each having eight coils assembled radially and symmetrically around the beam axis.
 
The inner-detector system (ID) is immersed in a \SI{2}{\tesla} axial magnetic field and provides charged-particle tracking in the range $|\eta| < 2.5$. A high-granularity silicon pixel detector covers the vertex region and usually provides four measurements per track. The pixel detector is followed by a silicon microstrip tracker which usually provides four measurement points per track. These silicon detectors are complemented by a transition radiation tracker, which enables radially extended track reconstruction and improved momentum measurements up to $|\eta| = 2.0$. The transition radiation tracker also provides electron identification information based on the fraction of hits (typically 30 hits in total) above a high-energy threshold designed for optimal electron--pion separation.
 
The calorimeter system covers the pseudorapidity range $|\eta| < 4.9$. Within the region $|\eta|< 3.2$,
electromagnetic calorimetry is provided by barrel and endcap lead/liquid-argon (LAr) sampling calorimeters,
with an additional thin LAr presampler covering $|\eta| < 1.8$, to correct for energy loss in material upstream of the calorimeters. Hadronic calorimetry is provided by a steel/scintillating-tile calorimeter, segmented into three barrel structures within $|\eta| < 1.7$, and two copper/LAr hadronic endcap calorimeters. The solid angle coverage is completed with forward copper/LAr and tungsten/LAr calorimeter modules
optimised for electromagnetic and hadronic measurements, respectively.
 
The muon spectrometer (MS) comprises separate trigger and high-precision tracking chambers measuring the deflection of muons in a magnetic field generated by superconducting air-core toroids. The field integral of the toroids ranges between \num{2.0} and \SI{6.0}{\tesla$\cdot$\metre} across most of the detector. A set of precision chambers covers the region $|\eta| < 2.7$ with three layers of monitored drift tubes, complemented by cathode strip chambers in the first measurement layer of the forward region,
where the background is highest. The muon trigger system covers the range $|\eta| < 2.4$ with resistive-plate chambers in the barrel, and thin-gap chambers in the endcap regions.
 
A two-level trigger system~\cite{TRIG-2016-01} is used to select events for offline analysis. The first-level trigger is implemented in hardware and uses a subset of the detector information. This is followed by the software-based high-level trigger, reducing the event rate to about 1 kHz.
 

\section{Data and simulation}
\label{sec:datamc}
 
This measurement utilises data collected by the ATLAS detector during the 2015 and 2016 data-taking periods. The data were recorded with a combination of single-lepton triggers, picking up events containing either an isolated lepton above a low-\pt~threshold or a high-\pt~lepton without any isolation requirement. The lower \pt~threshold for the isolated electron (muon) trigger ranges from 24 (20) to 26 \GeV~depending on the instantaneous luminosity. The higher \pt~threshold is 50 (60) \GeV~for the electron (muon) case over all the data-taking periods. Signal events satisfying the event selection described in \Sect{\ref{sec:sel}} are expected to have an overall trigger efficiency of 98\%.
 
Monte Carlo event simulation was deployed to model the signal and various background processes (summarised in \Tab{\ref{tab:simulation}}).
In the determination of integrated cross-sections, the \afactor{} and \cfactor{} factors as well as the \met{} shape for the $ZZ$ signal process were obtained from simulation.
The background contributions were either predicted by simulation or estimated in data with the assistance of simulation.
 
The $qqZZ$ process was modelled with \progname{Powheg-Box} v2~\cite{Nason:2004rx,Frixione:2007vw,Alioli:2010xd,Nason:2013ydw} interfaced to \progname{Pythia8.186}~\cite{Pythia8} for modelling of the parton showering, hadronisation and underlying event (UEPS).
The NLO matrix-element (ME) calculation set both the factorisation (\muf) and renormalisation (\mur) scales to the invariant mass of the $ZZ$ system (\mZZ), and used the NLO CT10~\cite{Lai:2010vv} parton distribution function (PDF). The UEPS
algorithm used a set of tuned parameters called the AZNLO tune~\cite{Aad:2014xaa}. The production cross-sections as a function of \mZZ{} were corrected to NNLO QCD and NLO EW accuracies in the total phase space.
The QCD $K$-factors were derived using the \progname{MATRIX} program~\cite{Grazzini:2015hta}, which computes the NNLO cross-section using the same QCD scales and the NNLO CT10 PDF~\cite{Gao:2013xoa}.
The EW correction was applied using $K$-factors provided by the authors of \Ref{\cite{NLOEWcorr}}.
The QCD correction is about +10\% for the entire \mZZ~spectrum, while the EW correction is about $-$4\% at low \mZZ~but has a larger impact at high \mZZ, which cancels out the positive QCD correction for \mZZ~around 500 \GeV. An alternative sample was generated with \progname{Sherpa2.2.2}~\cite{Gleisberg:2008ta} using the NNLO NNPDF3.0 PDF~\cite{Ball:2014uwa} and the same choice of QCD scales. The \progname{Sherpa} generator
and its associated UEPS algorithm has NLO QCD accuracy for inclusive observables and extended QCD precision for events with one or more outgoing partons (NLO for up to one parton, LO for two and three partons).
 
The $ggZZ$ events were simulated with the LO \progname{gg2vv3.1.6}~\cite{Kauer:2012hd,Kauer:2013qba} generator using the NNLO CT10 PDF, and then interfaced to \progname{Pythia8.186} using the A14 tune~\cite{ATL-PHYS-PUB-2014-021}. The production cross-section was corrected to NLO QCD accuracy using a $K$-factor of 1.7 reported in \Ref{\cite{Caola:2015psa}}.
An alternative modelling was provided by \progname{Sherpa2.1.1}~\cite{Cascioli:2013gfa} with the NLO CT10 PDF, which extended the LO QCD calculation to events with one parton. Both generators used \mZZ$/2$ for the QCD scales, and they incorporated both the fermion-loop and the Higgs processes, together with the interference between the two.
 
To study the effects of aTGCs, an additional sample for the SM $qqZZ$ process was generated at NLO in QCD using \progname{Sherpa2.1.1} with the NLO CT10 PDF. The simulated sample was interfaced to a parton-level program~\cite{Baur:2000ae} following the procedures detailed in Ref.~\cite{Bella:2008wc}, and then event-by-event weights reflecting the relative change in  the cross-sections due to any aTGCs were computed. A parameterisation of aTGC contributions as a function of any kinematic variable can be derived with this information. This procedure was adopted in the previous $ZZ$ measurements~\cite{STDM-2012-02,STDM-2014-16,STDM-2016-15}.
 
Production of \ZZllll{} events was modelled in the same way as the signal events.
The diboson background processes $WZ$ and $WW$ were generated with \progname{Powheg-Box} v2 using the NLO ME calculation and the NLO CT10 PDF, and then interfaced to \progname{Pythia8.186} with the AZNLO tune.
An alternative $WZ$ sample was produced with \progname{Powheg-Box} v2 interfaced to \progname{Herwig++}~\cite{Bahr:2008pv},
for the study of UEPS uncertainties.
The interference between the $WW$ and $ZZ$ processes in the \llnn{} final state was found to be negligible~\cite{Kallweit:2017khh} and was therefore not considered in this analysis.
Both the \ttbar{} and $Wt$ events were simulated at NLO in QCD with \progname{Powheg-Box} v2~\cite{Frixione:2007nw,Re:2010bp}
and interfaced to \progname{Pythia6.428}~\cite{Sjostrand:2006za}, and the production cross-sections were corrected to
NNLO QCD precision~\cite{Czakon:2011xx,Kidonakis:2010ux}.
\progname{Sherpa2.2.1} with the NNLO NNPDF3.0 PDF was used to model the \zjet{} process.
The production cross-section for the \zjet{} process was calculated with NNLO QCD precision,
while the simulation has NLO QCD precision for events with zero, one and two partons, and provided a LO QCD description for events with three to five partons.
The rare $VVV$ background, consisting of $WWW$, $WWZ$, $WZZ$ and $ZZZ$ processes, was modelled with \progname{Sherpa2.1.1} with NLO QCD precision. \progname{MadGraph5\_aMC@NLO}~\cite{Alwall:2014hca} interfaced to \progname{Pythia8.186} was used to generate the $t\bar{t}V$ background events that account for $\ttbar W$, $\ttbar Z$ and $\ttbar WW$ production processes. The $t\bar{t}V$ process was calculated at LO QCD accuracy, and its production cross-section was corrected to NLO QCD precision~\cite{Alwall:2014hca}.
 
Generated events were then processed through the ATLAS detector simulation~\cite{simuAtlas} based on GEANT4~\cite{Agostinelli:2002hh} to emulate the response of the detector to the final-state particles.
Pile-up was simulated with \progname{Pythia8.186} using the A2 tune~\cite{ATL-PHYS-PUB-2011-014} and overlaid on simulated events to mimic the real collision environment.
The distribution of the average number of interactions per bunch crossing in the simulation was weighted to reflect that in data.
Simulated events were processed with the same reconstruction algorithms as for the data.
Furthermore, the lepton momentum scale and resolution, and the lepton reconstruction, identification, isolation and trigger efficiencies in the simulation
were corrected to match those measured in data.
 
\begin{table}[t]
\begin{center}
\scalebox{0.85}{
\begin{tabular}{l c c c}
\hline
\hline \noalign{\smallskip}
Process & Generator & Simulation accuracy & Cross-section accuracy \\
\noalign{\smallskip}\hline \noalign{\smallskip}
\multirow{2}{*}{$qqZZ$}
& \progname{Powheg-Box} v2 + \progname{Pythia8.186} & NLO QCD & NNLO QCD + NLO EW \\ [5pt]
& \progname{Sherpa2.2.2} & \multicolumn{2}{c}{NLO QCD 0-1p, LO QCD 2-3p}\\ [5pt]
\multirow{2}{*}{$ggZZ$}
& \progname{gg2vv3.1.6} + \progname{Pythia8.186} & LO QCD & NLO QCD \\ [5pt]
& \progname{Sherpa2.1.1} & \multicolumn{2}{c}{LO QCD 0-1p} \\ [5pt]
$qqZZ$ (aTGCs) &  \progname{Sherpa2.1.1} & \multicolumn{2}{c}{NLO QCD 0-1p, LO QCD 2-3p} \\ [5pt]
\noalign{\smallskip}\hline \noalign{\smallskip}
\multirow{2}{*}{$WZ$}
& \progname{Powheg-Box} v2 + \progname{Pythia8.186} & \multicolumn{2}{c}{\multirow{2}{*}{NLO QCD}}\\ [5pt]
& \progname{Powheg-Box} v2 + \progname{Herwig++} &  \multicolumn{2}{c}{}\\ [5pt]
$WW$ & \progname{Powheg-Box} v2 + \progname{Pythia8.186} & \multicolumn{2}{c}{NLO QCD} \\ [5pt]
$qq$\ZZllll    & \progname{Powheg-Box} v2 + \progname{Pythia8.186} & NLO QCD & NNLO QCD + NLO EW \\ [5pt]
$gg$\ZZllll    & \progname{gg2vv3.1.6} + \progname{Pythia8.186} & LO QCD & NLO QCD \\ [5pt]
\zjet      & \progname{Sherpa2.2.1} & NLO QCD 0-2p, LO QCD 3-5p & NNLO QCD \\ [5pt]
\ttbar     & \progname{Powheg-Box} v2 + \progname{Pythia6.428}  & NLO QCD & NNLO QCD \\ [5pt]
$Wt$        & \progname{Powheg-Box} v2 + \progname{Pythia6.428}  & NLO QCD & NNLO QCD \\ [5pt]
$VVV$      & \progname{Sherpa2.1.1} & \multicolumn{2}{c}{NLO QCD} \\ [5pt]
$t\bar{t}V$ & \progname{MadGraph5\_aMC@NLO} + \progname{Pythia8.186} & LO QCD & NLO QCD \\
\noalign{\smallskip}\hline
\hline
\end{tabular}
}
\caption{Summary of Monte Carlo event simulation tools with their theoretical accuracy for each process, where ``p'' stands for parton(s).
For the first two signal processes and the $WZ$ process,
the first (second) row describes the
baseline (alternative) simulation. The theoretical accuracy of the normalisation used for the total production cross-section of each process is shown in the last column.}
\label{tab:simulation}
\end{center}
\end{table}

\section{Selection of \llnn~events}
\label{sec:sel}
 
This analysis selects a detector signature with a pair of high-\pT~isolated electrons ($ee$) or muons ($\mu\mu$) and large \met.
The $ee$ and $\mu\mu$ channels are combined to obtain the final results. The event selection strategy was optimised to cope with the large background contamination.
The selection requirements lead to a highly boosted $Z$ boson back-to-back with the missing transverse momentum vector (\vecmet). Backgrounds are further reduced by removing events with extra leptons or any jets containing $b$-hadrons (``$b$-jets''). Therefore, a precise understanding of the overall reconstruction and selection of leptons, jets, and \met{} is required in this measurement.
 
Events are first required to have a collision vertex associated with at least two tracks each with $\pt>0.4$ \GeV.
The vertex with the highest scalar $p_{\mathrm{T}}^{2}$ sum of the associated tracks is referred to as the primary vertex.
 
Electrons are reconstructed from energy deposits in the EM calorimeter matched to a track reconstructed in the ID.
The electron identification imposes selections on the number of hits in the ID
and requirements on a likelihood discriminant, built from variables related to EM calorimeter shower shapes, track--cluster matching, track quality, and transition radiation~\cite{Aaboud:2019ynx}. Electrons must satisfy the ``medium'' identification criterion~\cite{Aaboud:2019ynx}, which is about 90\% efficient for electrons with \pT $\approx$ 40 \GeV. Candidate electrons must have $\pT>7$~\GeV{} and pseudorapidity $|\eta|<2.47$.
Muons are reconstructed by combining all the hits associated with a pair of matched tracks reconstructed in the ID and MS,
taking into account the energy loss in the calorimeter. Muons are identified by requiring a sufficient number of ID and MS hits, and good consistency between the ID and MS track measurements as well as good combined fit quality~\cite{Aad:2016jkr}, and they must satisfy the ``medium'' identification criterion~\cite{Aad:2016jkr}, which has an overall efficiency of 96\%.
Candidate muons are required to have $\pT>7$~\GeV{} and $|\eta|<2.5$.
To further suppress misidentified lepton and cosmic-ray background contributions, the absolute value of the longitudinal impact parameter of leptons with respect to the primary vertex   must be smaller than 0.5 mm, and the transverse impact parameter divided by its error must be less than 5 (3) for electrons (muons). In addition, the ``loose'' isolation criteria defined in Refs.~\cite{Aad:2016jkr,Aaboud:2019ynx}~are applied. The isolation selection imposes requirements on both the track-based and calorimeter-based isolation variables, and varies as a function of \pT to maintain a uniform efficiency above 98\% for prompt leptons.
 
Jets are reconstructed with the anti-$k_t$ algorithm~\cite{antikt} with radius parameter $R = 0.4$, using as input positive-energy topological clusters in the calorimeters~\cite{jets,Aaboud:2017jcu,Cacciari:2011ma}. The jet energy scale is calibrated using simulation and further corrected with in situ methods~\cite{Aaboud:2017jcu}.
Candidate jets must have $\pT>20$~\GeV{} and $|\eta|<4.5$. Additional requirements using the track and vertex information inside a jet~\cite{jvt} are applied for jets with $\pT<60$~\GeV{} and $|\eta|<2.5$ to suppress pile-up contributions.
Candidate $b$-jets ($\pt > 20$~\GeV{} and $|\eta| <  2.5$) are identified with an algorithm providing 85\% signal efficiency and a rejection factor of 33 against light-flavour jets~\cite{bjets-13TeV}.
 
Leptons and jets may be close to each other or overlapping, even after implementing the full set of object selections.
The appearance of such overlapping objects may lead to ambiguities in the event selection and in the energy measurement of the physics objects.
A common procedure in the ATLAS experiment, as detailed in \Ref{\cite{Aad:2016eki}}, is applied to resolve the ambiguities.
This requirement helps to suppress the occurrence of two ``problematic'' scenarios, one with energy measurement of electrons biased due to nearby jets, and the other with a jet producing non-prompt muons through meson decays in flight.

The \vecmet{} vector is computed as the negative of the vector sum of transverse momenta of all the leptons and jets, as well as the tracks originating from the primary vertex but not associated with any of the leptons or jets (``soft-term'')~\cite{ATL-PHYS-PUB-2015-027}.
The soft-term is computed in a way minimising the impact of pile-up in the \met{} reconstruction.
 
Candidate events are preselected by requiring exactly two selected electrons or muons with opposite charges and $\pT>20$~\GeV.
The leading lepton is further required to have $\pT>30$~\GeV,
well above the threshold of the single-lepton triggers.
To suppress the $WZ$ background, events containing any additional
lepton satisfying the ``loose'' rather than ``medium'' identification requirement, in addition to
the other requirements, are rejected.
The dilepton invariant mass ($m_{\ell\ell}$) is required to be in the range between 76 and 106~\GeV,
which largely reduces the contamination from the \nonresll{} background.
\Fig{\ref{fig:met_pre}} shows the observed and expected \met{} spectra after imposing the above requirements (``preselection'').
The fractional experimental uncertainties in the expected spectra increase as a function of \met{} in the region dominated
by the \zjet{} process, as a result of the asymmetric migration effects along the steeply falling \met{} distribution and the large
jet-related uncertainty for \zjet{} events at high \met{}. The top-quark processes with genuine \met{} dominate the high \met{} region.
For \met{} around 200 \GeV, top-quark events generally contain less jet activity than \zjet{} events:
this leads to correspondingly smaller experimental uncertainties.
The experimental uncertainties are elaborated in \Sect{\ref{sec:unc}}.
The data sample after the preselection is dominated by the
\zjet{} and \nonresll{} processes. To suppress these backgrounds, a further selection based on \met{} and event topology is applied.
 
Candidate events are required to have $\met>110$~\GeV{} and \VToST~$>0.65$, where \VT{} is the magnitude of the vector sum of transverse momenta of selected leptons and jets, and \ST{} is the scalar \pT{} sum of the corresponding objects.
The variable \VToST{} was found to be less sensitive to jet experimental uncertainties than similar variables such as \METoST.
To further reduce the impact of jet energy scale uncertainties, the calculation of \VT{} and
\ST{} uses ``hard jets'' which are required to have $\pT>25$~\GeV{} for the central region ($|\eta|<2.4$)
and $\pT>40$~\GeV{} for the forward region ($2.4 < |\eta| < 4.5$).
The \met{} cut suppresses the \zjet{} contamination by many orders of magnitude,
and the residual \zjet{} events, which have large fake \met, are further suppressed by the \VToST{} requirement.
As the consequence of the combined \met{} and \VToST{} requirement, the \zjet{} process only constitutes a small fraction of the total background after the full selection.

\begin{figure}[!t]
\begin{center}
\includegraphics[width=0.495\textwidth]{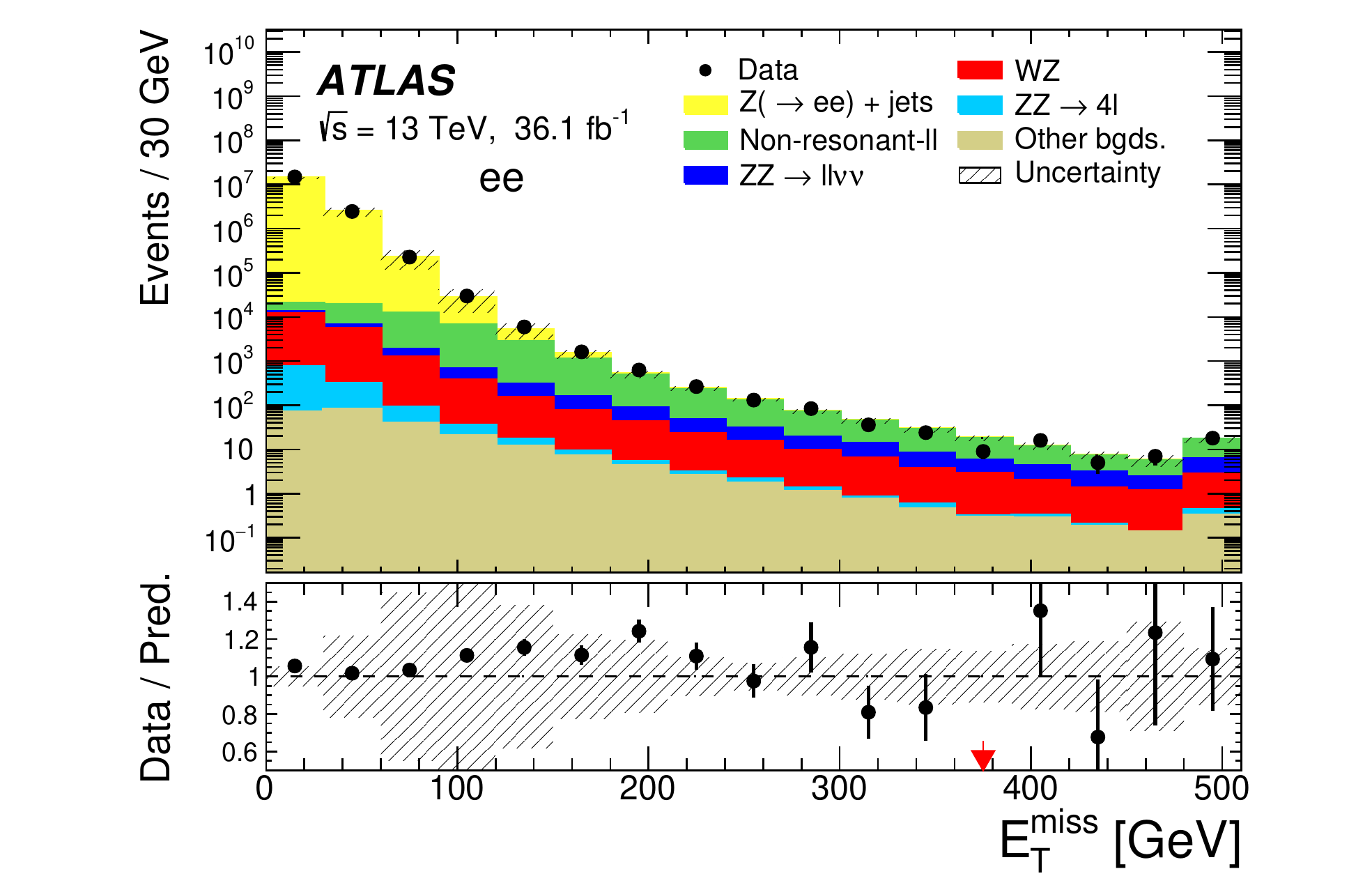}
\includegraphics[width=0.495\textwidth]{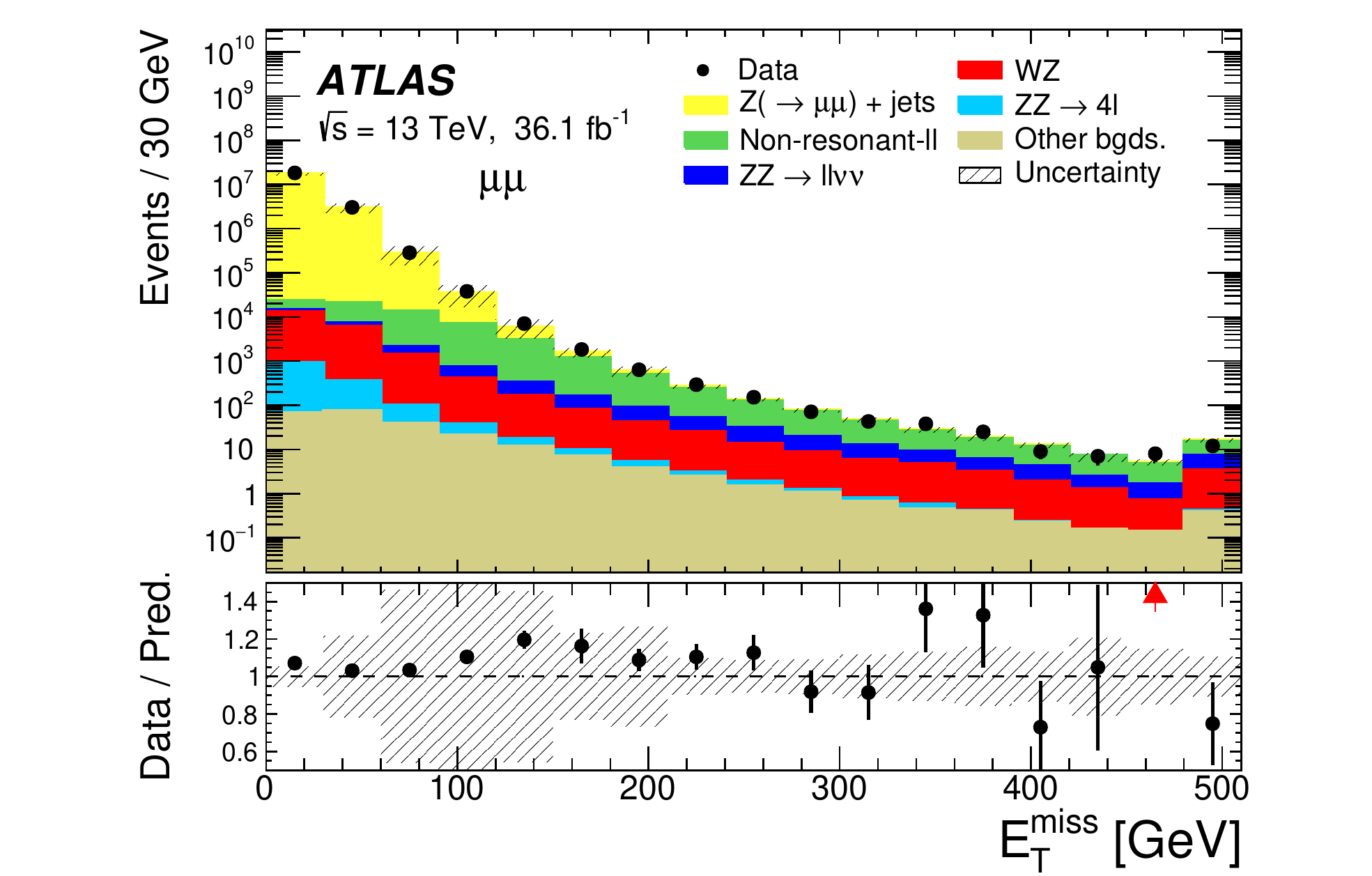}
\end{center}
\caption{The \met{} distributions after the preselection for the $ee$ (left) and $\mu\mu$ (right) channels.
The expectation is derived from simulation and the hashed band accounts for the statistical and
experimental uncertainties. The experimental uncertainties are described in \Sect{\ref{sec:unc}}.
The last bin in the distributions contains the overflow events. The arrows indicate that the differences
between the data and the expectation in some bins exceed the $y$-axis scope of the bottom plots.
The ``other'' background corresponds to the $VVV$ and $t\bar{t}V$ processes.
}
\label{fig:met_pre}
\end{figure}

\begin{table}[t]
\begin{center}
\begin{tabular}{l c}
\hline
\hline\noalign{\smallskip}
Step                &      Selection criteria \\
\noalign{\smallskip}\hline \noalign{\smallskip}
Two leptons         & Two opposite-sign leptons, leading (subleading) $\pt > 30$ $(20)$ \GeV \\
\noalign{\smallskip}\hline \noalign{\smallskip}
Jets                & \pT~$>$ 20 \GeV{}, $|\eta|<$ 4.5, and $\Delta R > 0.4$ relative to the leptons \\
\noalign{\smallskip}\hline \noalign{\smallskip}
Third-lepton veto   & No additional lepton with $\pt > 7$ \GeV \\
\noalign{\smallskip}\hline \noalign{\smallskip}
$m_{\ell\ell}$      & 76 $<m_{\ell\ell}<$ 106 \GeV \\
\noalign{\smallskip}\hline \noalign{\smallskip}
Hard jets    &  $\pT>25$~\GeV{} for $|\eta|<2.4$, $\pT>40$~\GeV{} for $2.4 < |\eta| < 4.5$ \\
\noalign{\smallskip}\hline \noalign{\smallskip}
\met and \VToST     & \met~$>$ 110 \GeV{} and \VToST~$>$ 0.65  \\
\noalign{\smallskip}\hline \noalign{\smallskip}
\drll              &  \drll~$<$ 1.9 \\
\noalign{\smallskip}\hline \noalign{\smallskip}
\dphizmet          &  \dphizmet~$>$ 2.2 radians \\
\noalign{\smallskip}\hline \noalign{\smallskip}
$b$-jet veto & $N$($b$-jets) $= 0$ with $b$-jet \pt~$>$ 20 \GeV~and $|\eta|<$ 2.5 \\
\noalign{\smallskip}\hline
\hline
\end{tabular}
\end{center}
\caption{Event selection criteria for the \llnn~signature.}
\label{tab:selection}
\end{table}

Additional selection criteria based on angular variables are imposed to ensure the desired detector signature, which helps to further reject the \zjet~and \nonresll~background events. The azimuthal angle difference between the dilepton system and \vecmet, \dphizmet, must be larger than 2.2 radians, and the selected leptons must be close to each other, with the distance \drll~$=\sqrt{(\Delta\phi_{\ell\ell})^2 + (\Delta\eta_{\ell\ell})^2}<$ 1.9.
Finally, events containing one or more $b$-jets are vetoed to further suppress the \ttbar~and $Wt$ backgrounds. The full event selection is summarised in \Tab{\ref{tab:selection}}. \Fig{\ref{fig:sequential_cuts}} gives the observed and simulated spectra for \VToST, \drll, \dphizmet, and the number of $b$-jets, where each plot is made with the implementation of all the cuts prior to the cut on that variable, according to the cut sequence in Table~\ref{tab:selection}.
 
\begin{figure}[!ht]
\begin{center}
\includegraphics[width=0.495\textwidth]{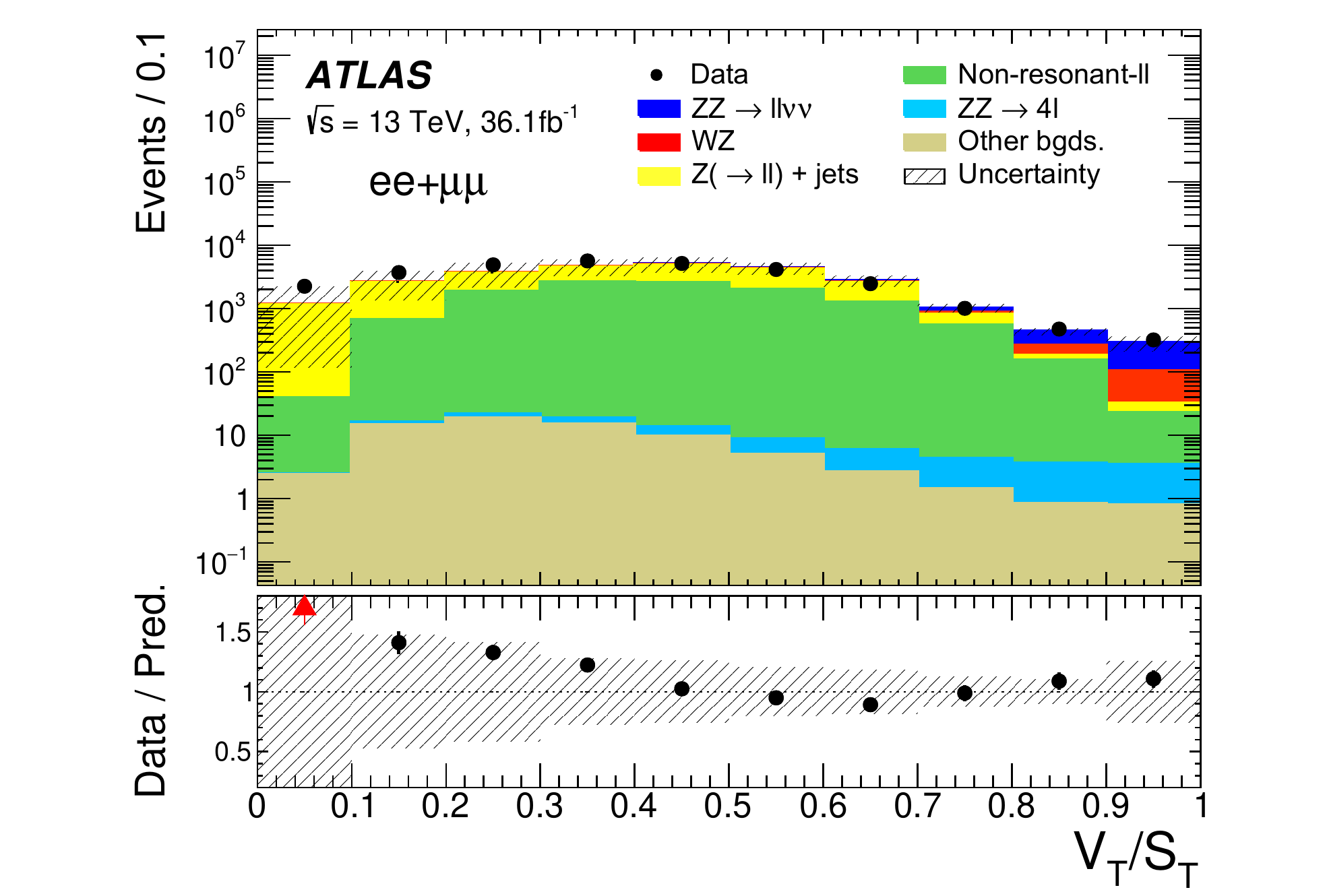}
\includegraphics[width=0.495\textwidth]{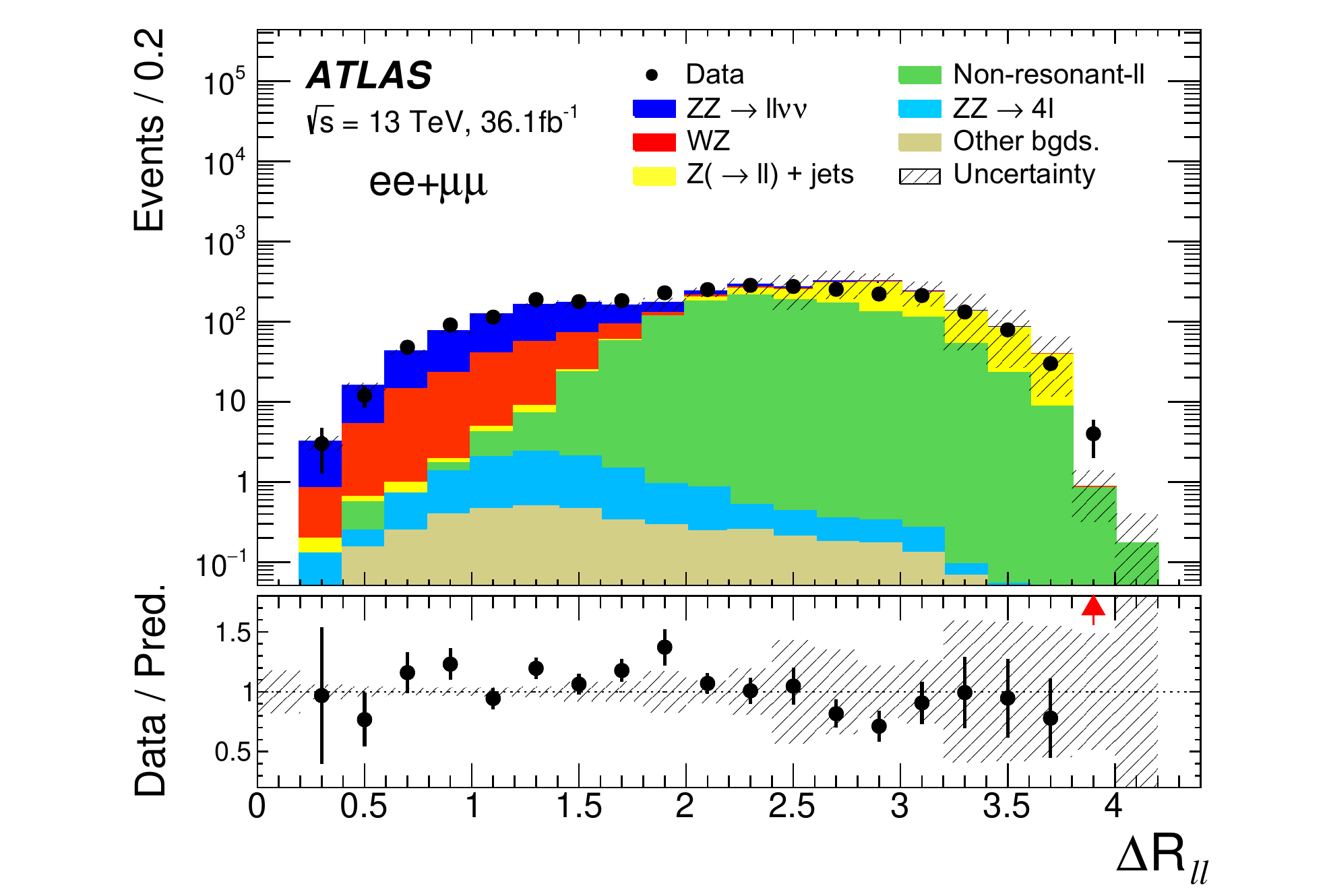}
\includegraphics[width=0.495\textwidth]{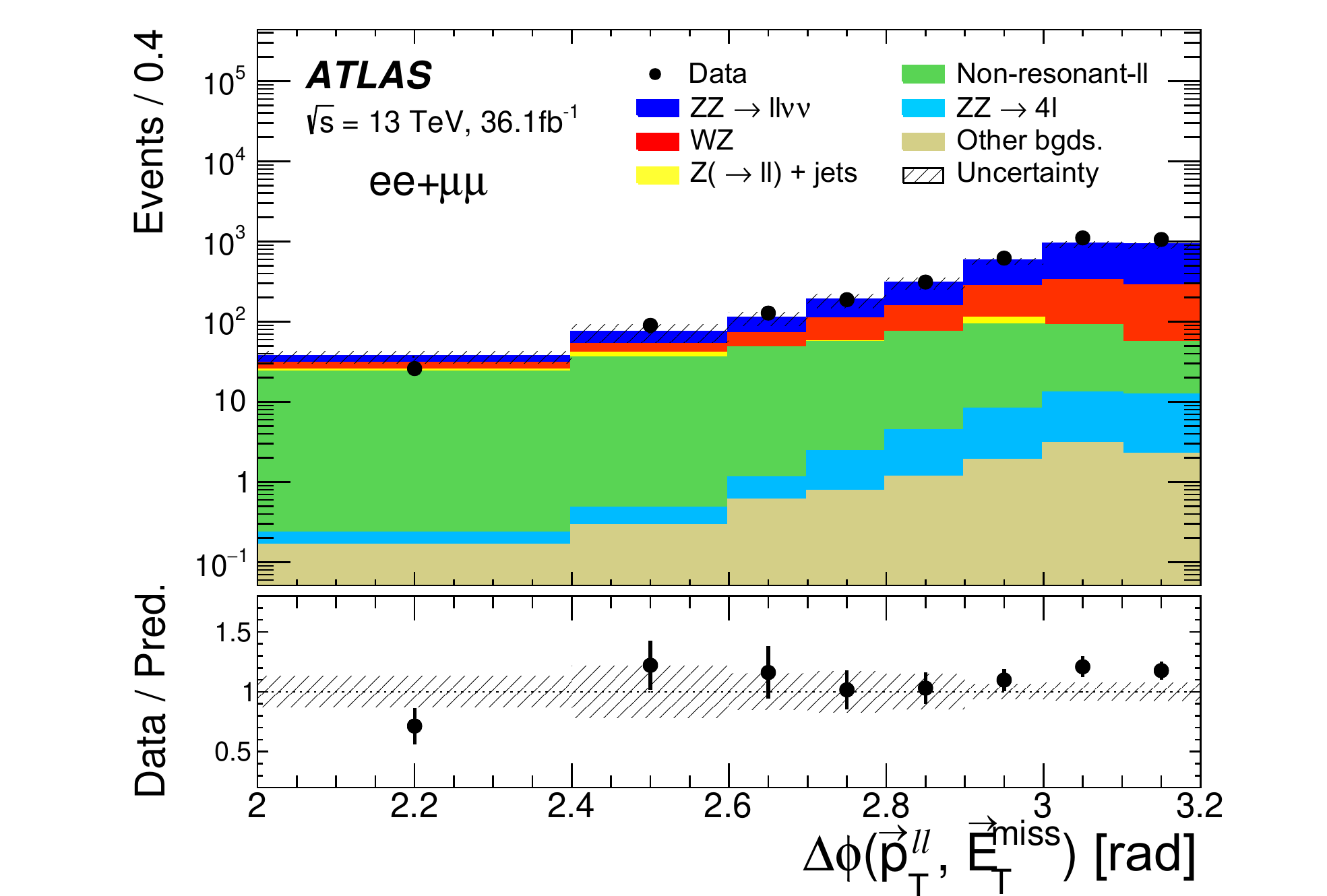}
\includegraphics[width=0.495\textwidth]{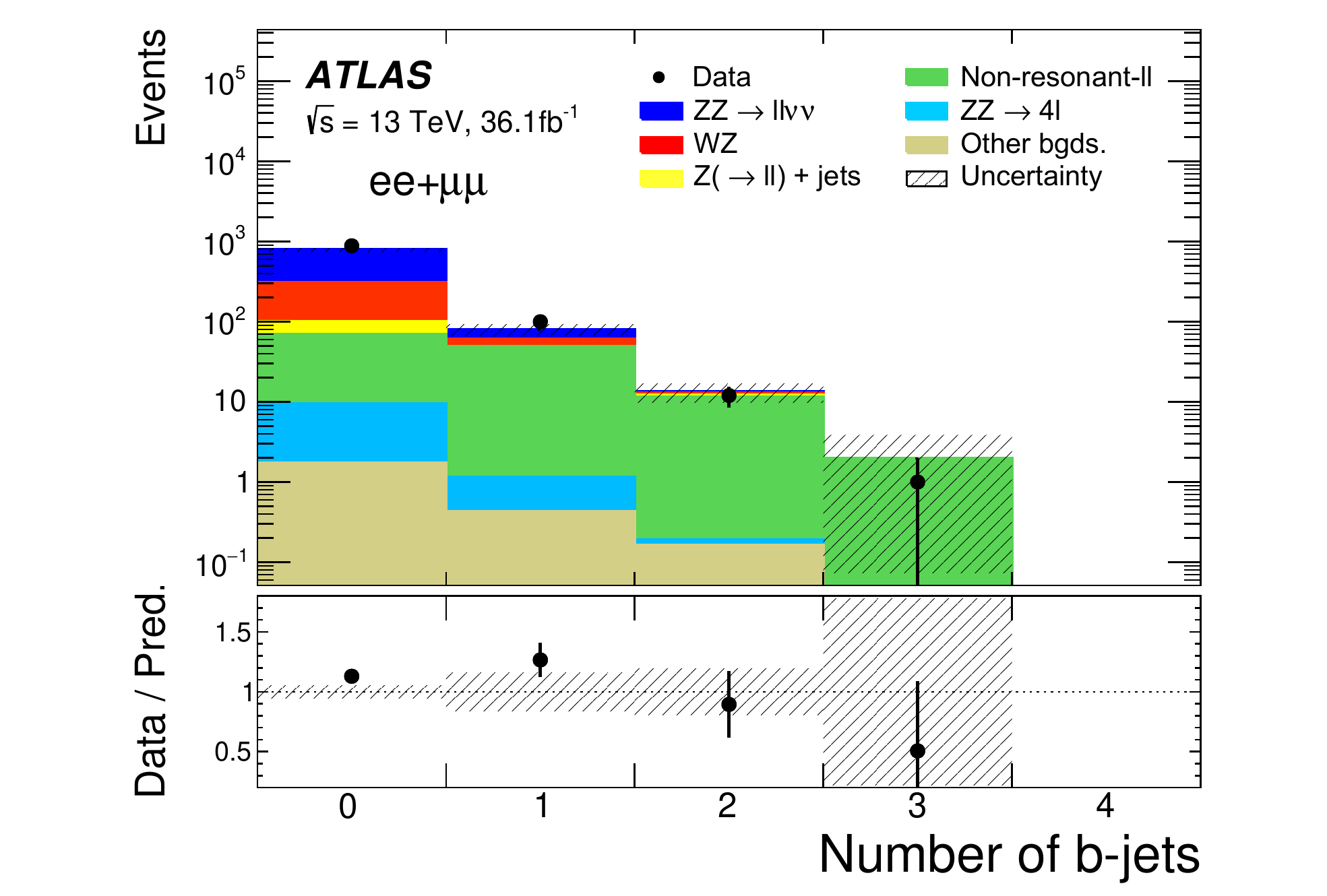}
\caption{The \VToST, \drll, \dphizmet, and the number of $b$-jets distributions for the combination of $ee$ and $\mu\mu$ channels
with the implementation of all the cuts in \Tab{\ref{tab:selection}} prior to the cut on that variable.
The expectation is derived from simulation and the hashed band accounts for
the statistical and experimental uncertainties.
The first bin in the distribution of \dphizmet{} (bottom left) contains the underflow events.
The arrow in the \VToST{} distribution indicates that the difference between the data and the expectation
exceeds the $y$-axis scope of the bottom plot.
The ``other'' background corresponds to the $VVV$ and $t\bar{t}V$ processes.
}
\label{fig:sequential_cuts}
\end{center}
\end{figure}
 

\section{Total and fiducial phase spaces}
\label{sec:ps}
 
The definitions of the total and fiducial phase spaces are summarised in \Tab{\ref{tab:ps}}.
The total phase space is defined as in \Ref{\cite{STDM-2016-15}} for the $ZZ\to 4\ell$ measurement,
requiring $66 < m_{\ell\ell}, m_{\nu\nu} < 116$ \GeV~($\ell = e$ or $\mu$), where the leptons and neutrinos originate from the $Z$-boson decays.
The four-momenta of the leptons are defined at Born level, i.e. before any QED final-state radiation.

The fiducial phase space is defined with a set of criteria very close to that of the detector-level event selection (\Tab{\ref{tab:selection}}).
This strategy helps to reduce the amount of phase-space extrapolation in the fiducial measurements and therefore minimises the theoretical uncertainties of the results. The criteria are applied to ``particle-level'' physics objects, which are reconstructed from stable final-state particles, prior to their interactions with the detector.
For electrons and muons, QED final-state radiation is partly recovered by adding to the lepton four-momentum
the four-momenta of surrounding photons not originating from hadrons within an angular distance $\Delta R < 0.1$
(dressed leptons).
Particle-level jets are built with the anti-$k_t$ algorithm with radius parameter $R = 0.4$, using all final-state particles as the input (excluding muons and neutrinos).
As shown in \Tab{\ref{tab:ps}}, the selection requirements for the fiducial phase space closely follow those in \Tab{\ref{tab:selection}}.
The \vecmet~vector is defined as the sum of transverse momenta of the two neutrinos from the $Z$-boson decays. This measurement requires large \met, which has a detector resolution of around 10 \GeV~\cite{ATL-PHYS-PUB-2015-027} in the phase space considered here. To accommodate the majority of the events selected at detector level, the \met~threshold is therefore lowered to 90 \GeV~in the fiducial region.
The efficiency of the $b$-jet veto is found to be 98\% in the fiducial region and consistent between the \progname{Powheg+Pythia8} and \progname{Sherpa} generators. No requirement is made on the number of $b$-jets in the fiducial selection.

\Tab{\ref{tab:psnum}} gives the expected signal yields at detector level, the \afactor~and \cfactor~factors, and the predicted cross-sections. The $qqZZ$ and $ggZZ$ processes have similar final-state kinematic distributions
and their \afactor{} and \cfactor{} factors are similar.
The corresponding factors for the total $ZZ$ process are averaged from that for the two sub-processes, weighted by the respective cross-sections. The cross-section predictions for the total phase space are corrected for
the branching fraction of the \ZZllnn{} decays, 1.35\% with a negligible uncertainty, obtained from Refs.~\cite{PDG2018,LEP_SLD}.
The expected fiducial and total cross-sections, \xsfidexp{} and \xstotexp, are calculated from simulation, including the higher-order corrections detailed in \Sect{\ref{sec:datamc}}. The total uncertainties in these predictions are also provided in \Tab{\ref{tab:psnum}}, and the procedures used to derive these uncertainties are
described in \Sect{\ref{sec:unc}}.
 
\begin{table}[!t]
\begin{center}
 
\renewcommand{\arraystretch}{1.2}
\begin{tabular}{l | p{10cm}}
\hline
\hline
\multirow{2}{*}{Total phase space}
& Born-level leptons ($ee$ or $\mu\mu$) \\ 
& 66 < $m_{\ell\ell}, m_{\nu\nu} < 116~\GeV $\\
\hline
\multirow{6}{*}{Fiducial phase space}
& Dressed leptons ($e$ or $\mu$): \pt $>$ 7 \GeV, $|\eta| <$ 2.5 \\
& Jets: \pt $>$ 20 \GeV, $|\eta| <$ 4.5 \\
& Reject leptons if overlapping with a jet within $\Delta R <$ 0.4 \\
& Two leptons with leading (subleading) \pt $>$ 30 (20) \GeV \\
& 76 $<m_{\ell\ell}<$ 106 \GeV \\
& \met~$>$ 90 \GeV{} and \VToST~$>$ 0.65 \\
& \dphizmet~$>$ 2.2 radians and \drll~$<$ 1.9 \\
\hline
\hline
\end{tabular}
 
\caption{Definitions of the total and fiducial phase spaces for the $ZZ\to \ell\ell\nu\nu$ signal.}
\label{tab:ps}
\end{center}
\end{table}

\begin{table}[!htbp]
\begin{center}
\scalebox{0.85}{
\renewcommand{\arraystretch}{1.5}
\begin{tabular}{l | c  c | c  c | c  c }
\hline
\hline
& \multicolumn{2}{c|}{$ZZ$} & \multicolumn{2}{c|}{$qqZZ$} & \multicolumn{2}{c}{$ggZZ$} \\
\hline
& \multicolumn{1}{c}{$ee$} & $\mu\mu$ & \multicolumn{1}{c}{$ee$} & $\mu\mu$ & \multicolumn{1}{c}{$ee$} & $\mu\mu$ \\
\hline
 
\multirow{3}{*}{Signal yield}
&  \numRF{219.5}{3} $\pm$ \numRF{14.6}{2} & \numRF{228.8}{3} $\pm$ \numRF{15.3}{2} & & & & \\
&  ($\pm$ \numRF{2.2}{1} $_{\mathrm{stat}}$ $\pm$ \numRF{6.5}{1} $_{\mathrm{exp}}$   &  ($\pm$ \numRF{2.3}{1} $_{\mathrm{stat}}$ $\pm$ \numRF{6.9}{1} $_{\mathrm{exp}}$
& \numRF{194.4}{3} $\pm$ \numRF{12.0}{2} & \numRF{202.4}{3} $\pm$ \numRF{12.4}{2} & \numRF{25.1}{2} $\pm$ \numRF{15.1}{2} & \numRF{26.4}{2} $\pm$ \numRF{15.9}{2} \\
&   $\pm$ \numRF{12.9}{2} $_{\mathrm{theory}}$)     &  $\pm$ \numRF{13.4}{2} $_{\mathrm{theory}}$)                & & & & \\
 
\hline
 
\cfactor   & \multicolumn{2}{c|}{-} & (\numRF{54.7}{3} $\pm$ \numRF{1.7}{2})\% & (\numRF{56.6}{3} $\pm$ \numRF{1.7}{2})\% & (\numRF{53.1}{3} $\pm$ \numRF{1.8}{2})\% & (\numRF{55.5}{3} $\pm$ \numRF{2.2}{2})\% \\
\xsfidexp
& \multicolumn{2}{c|}{22.4 $\pm$ 1.3 fb} & \multicolumn{2}{c|}{18.8 $\pm$ 1.0 fb} & \multicolumn{2}{c}{2.6 $\pm$ 0.8 fb}\\
\afactor   & \multicolumn{2}{c|}{-} & \multicolumn{2}{c|}{(5.3 $\pm$ 0.1)\%} & \multicolumn{2}{c}{(5.3 $\pm$ 0.3)\%} \\
\xstotexp
& \multicolumn{2}{c|}{15.7 $\pm$ 0.7 pb} & \multicolumn{2}{c|}{13.9 $\pm$ 0.4 pb} & \multicolumn{2}{c}{1.8 $\pm$ 0.6 pb}\\
\hline
\hline
\end{tabular}
\renewcommand{\arraystretch}{1}
}
\caption{Predictions for the signal yields at detector level, for the \cfactor{} and \afactor{}
coefficients defined in Eq.~(\ref{eq:xs}), and for the cross-sections in the fiducial and
total phase spaces. The first column gives the corresponding predictions for the total $ZZ$ process, combined from
those shown separately for the $qqZZ$ and $ggZZ$ sub-processes.
The errors include both the statistical and systematic uncertainties (see Section~\ref{sec:unc}).
The statistical, experimental, and theoretical uncertainties are also shown separately
for the combined signal yields.
}
\label{tab:psnum}
\end{center}
\end{table}

 
\section{Background estimation}
\label{sec:bkg}
 
After the event selection, the overall signal-to-background ratio is about 1.7. The $WZ$ and \nonresll{} backgrounds account for 72\% and 21\% of the total background contribution, respectively, and are estimated from    control regions in data.
The \zjet{} background is largely suppressed, yielding a relative contribution of only 4\%, and is estimated from data.
Finally, the small remaining contributions from other processes, amounting in total to 3\% of the total background,
are estimated from simulation.
The various background estimates and their uncertainties are described below.
 
To estimate the dominant resonant background from $WZ$ production, a control region enriched in $WZ$ events, with a purity of 90\%,
is defined using the preselection criteria,
except that a third lepton with $\pt>20$~\GeV{} and satisfying the medium identification criteria is required.
Several further selections such as \VToST~$>0.3$, $b$-jets veto, and \mTW~$>60$~\GeV,
where \mTW{} is constructed from the third lepton's transverse momentum and the \vecmet{} vector,\footnote{$m_{\mathrm{T}}^{W} = \sqrt{2 \pT^{\ell} \met [1-\cos\Delta\phi({\vec{p}_{\mathrm{T}}^{\ell}}, \vec{E}_{\mathrm{T}}^{\mathrm{miss}})]}$}
are applied to suppress non-$WZ$ contributions.
A normalisation factor (\fWZ) is calculated in the control region as the number of observed events in data, subtracting the non-$WZ$ contributions estimated from simulation, divided by the predicted $WZ$ yield.
The factor \fWZ{} is found to be $1.26\pm0.04$ (stat),
which is consistent with a recent $WZ$ measurement~\cite{Aaboud:2019gxl}, performed within a broader fiducial phase space and
using a recent calculation of the $WZ$ total cross-section at NNLO in QCD~\cite{Grazzini:2016swo,Massimiliano:2017WZ}.
The statistical uncertainty of the data in the control region leads to a 3\% uncertainty in the $WZ$ estimate in the signal region.
The systematic uncertainty is evaluated for the ratio of the $WZ$ predictions in the signal and control regions. The experimental uncertainty in this ratio is 3.5\%, and the theoretical uncertainty is 3.3\%, calculated as the sum in quadrature of the PDF, scale, and UEPS uncertainties.
The non-$WZ$ contribution in the control region is less than 10\%. The uncertainty related to the subtraction of the non-$WZ$ contribution,
estimated by imposing cross-section uncertainties for all the relevant processes, is found to be about 2\%.
The total uncertainty in the $WZ$ estimate is about 6\%. The kinematic distributions are estimated from simulation, with both the experimental and theoretical uncertainties considered.
 
To estimate the \nonresll~background, including $WW$, top-quark (\ttbar~and $Wt$), and \Ztt{} production,
a control region dominated by the \nonresll~processes (with a purity above 95\%) is defined with all the event selection criteria
in \Tab{\ref{tab:selection}}, except that the final state is required to contain an opposite-sign $e\mu$ pair.
The \nonresll~contribution in the $ee$ ($\mu\mu$) channel is calculated as one half of the observed data yield after subtracting the contribution from the other background processes in the control region, and then corrected for the difference in the lepton reconstruction and identification efficiencies between selecting an $e\mu$ pair and an $ee$ ($\mu\mu$) pair.
The lepton efficiency correction is derived as the square root of the ratio of the numbers of $\mu\mu$ and $ee$ events in data after the preselection.
The choice of deriving the correction after preselection minimises the resulting statistical uncertainty.
The total uncertainty in the \nonresll~estimate is about 16\%, including the statistical uncertainty of the data in the control region (14\%) and the method bias estimated from simulation (7\%).
The kinematic distributions for the \nonresll~background estimate in the signal region are predicted with simulation,
and the assigned systematic uncertainty covers the experimental uncertainty in the simulated shape as well as the difference between data and simulation in the control region.
 
\Fig{\ref{fig:bkg_CR}} gives two examples of comparing data and predictions in the $WZ$ and \nonresll{} background control regions.
The left-hand figure is the \mTW{} distribution in the $WZ$ control region, where the normalisation factor \fWZ{} is applied to the $WZ$ simulation
and good agreement between the observed and predicted shapes is found. The right-hand figure is the \met{} distribution in the \nonresll{} control region, where
the $WW$ and top-quark (\ttbar{} and $Wt$) production processes are dominant. Both the statistical and experimental uncertainties are included in the
hashed bands in these figures.
 
\begin{figure}[t]
\begin{center}
\includegraphics[width=0.495\textwidth]{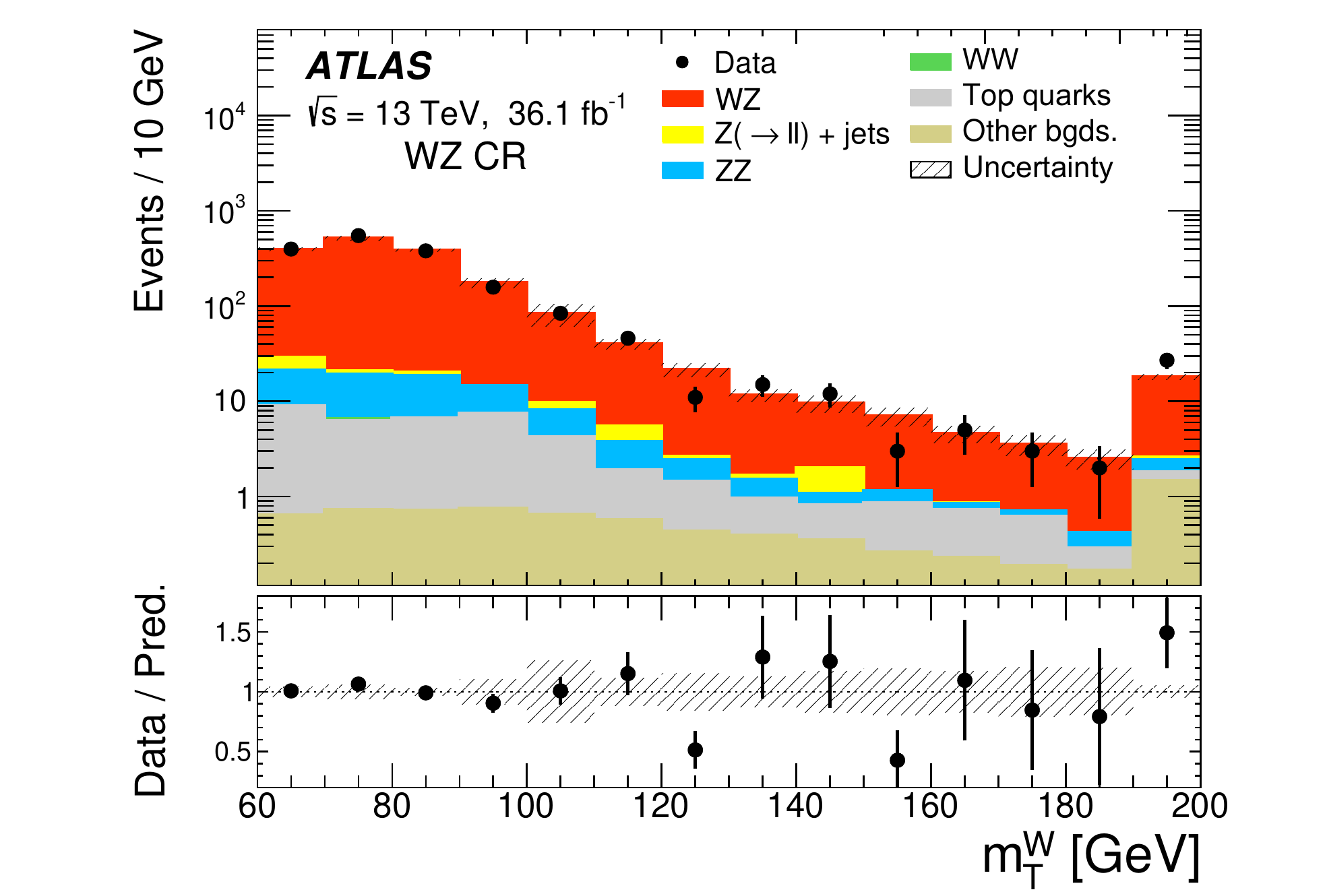}
\includegraphics[width=0.495\textwidth]{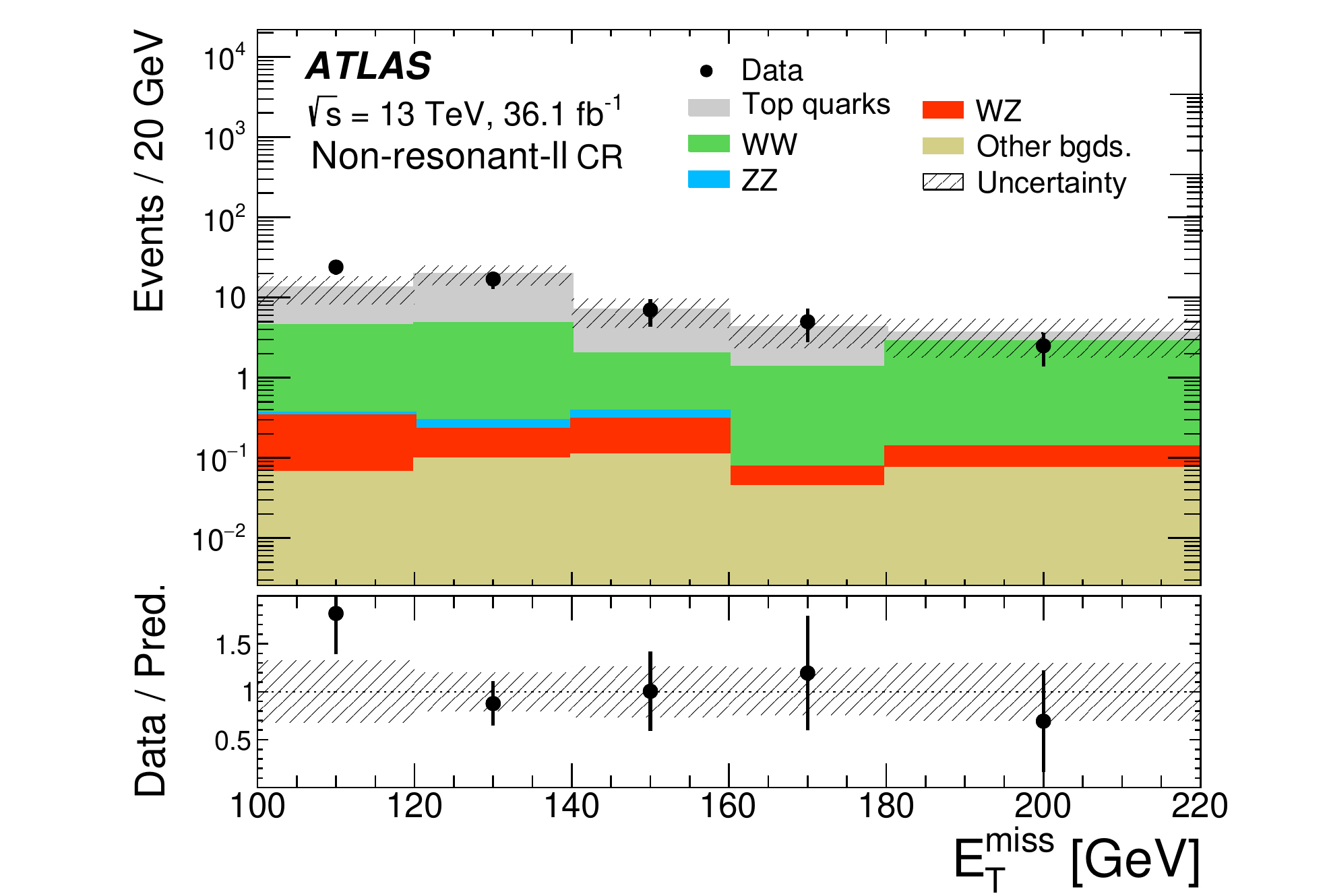}
\end{center}
\caption{Distributions in the control regions (CR), of \mTW{} for the $WZ$ CR (left) and
of \met{} for the \nonresll{} CR (right).
The data are compared with the predictions from simulation,
where the $WZ$ contribution is scaled by the normalisation factor of $1.26$
described in the text.
The last bin in the distributions contains the overflow events.
The hashed bands include both the statistical and experimental uncertainties.
The ``other'' background corresponds to the $VVV$ and $t\bar{t}V$ processes.
}
\label{fig:bkg_CR}
\end{figure}

A data-driven method is used to estimate the \zjet~background. This method defines three independent regions
(labelled as B, C and D)
which are enriched in \zjet~events and are not overlapping with the signal region (labelled as A). The data yields after subtracting the non-$Z$ contributions in these regions (\nB, \nC~and \nD) are used to predict the \zjet~contribution in the signal region (\nA), calculated as \BCD.
The main assumption of the method is that \eqABCD.
The control region definitions are optimised to ensure that this assumption is valid within uncertainties evaluated from simulation.
The control regions are defined using the preselection requirements plus the $b$-jets veto.
A further requirement of \met~$>30$~\GeV{} and \VToST~$>0.2$ is imposed to remove the low-\met{} phase space which is far away from the signal region.
The \met~and \VToST~variables are expected to have a small correlation with the topological variables,
so the various requirements to define the control regions are
grouped together, such that the correlations between regions are minimised. Specifically,
two Boolean variables are defined as, $\alpha$~$=$ ``\met~$>110$~\GeV{} and \VToST~$>0.65$'' and $\beta$~$=$ ``\dphizmet~$>2.2$~radians and \drll~$<1.9$''. The four regions are then defined as follows:
\begin{itemize}
\item Region A: \, $\alpha$ = TRUE \, \, and \, $\beta$ = TRUE
\item Region B: \, $\alpha$ = FALSE \, and \, $\beta$ = TRUE
\item Region C: \, $\alpha$ = TRUE \, \, and \, $\beta$ = FALSE
\item Region D: \, $\alpha$ = FALSE \, and \, $\beta$ = FALSE
\end{itemize}

Regions~B and D are dominated by the \zjet~process (with a purity greater than 95\%), while its relative contribution in region~C is
only 70\% because the \ttbar{} contribution in this phase space region remains large.
The derived \zjet{} contribution is corrected for the closure factor (\cfABCD) estimated from simulation. This factor is found to be 0.9 and has a relative uncertainty of 48\%, consisting of the statistical (40\%), experimental (22\%), and methodology uncertainties (15\%). The experimental uncertainty in the closure factor is dominated by jet energy scale and resolution.
The methodology uncertainty covers the variations obtained by changing the \met~and \VToST~thresholds in the low-\met~removal requirement
by 40\%.
The \zjet{} estimation is also subject to the statistical uncertainty of the data (5\%) and the subtraction of non-$Z$ contributions in the control regions (5\%).
The non-$Z$ subtraction uncertainty is driven by the modelling uncertainty for the main non-$Z$ process in region~C
(\ttbar{} production), which is about 10--20\% for \met{} above 100~\GeV~\cite{TOPQ-2015-02}.
The total uncertainty on the \zjet{} estimate is about 50\%.
The kinematic distributions for the \zjet{} background in the signal region are derived from the data in region~C,
together with a systematic uncertainty assigned in a way similar to that described above for the \nonresll{} background.

The \ZZllll{}, $VVV$ and $t\bar{t}V(V)$ backgrounds are estimated from simulation, and their contributions have a total uncertainty of
10--20\%, including both the theoretical cross-section~\cite{STDM-2016-15,Maltoni:2015ena,Binoth:2008kt} and the experimental uncertainties.

 
\section{Systematic uncertainties}
\label{sec:unc}
 
The measurement results and predictions are subject to theoretical and experimental uncertainties,
as well as uncertainties related to the background estimation.
The background uncertainties are explained in \Sect{\ref{sec:bkg}}.
The statistical uncertainties of the simulated samples for both the signal and background processes are
also taken into account wherever applicable. The systematic uncertainty sources for the signal process are detailed below.
 
The theoretical uncertainties for the dominant $qqZZ$ signal sub-process are estimated with the \\
\progname{Powheg+Pythia8} generator, since only the total cross-section has been calculated to NNLO QCD and NLO EW accuracies.  The theoretical uncertainties originate from the PDF choice, the missing higher-order QCD calculation, and the UEPS modelling. The PDF uncertainty is calculated as the 68\% confidence-level eigenvector uncertainty~\cite{Lai:2010vv} of the nominal PDF used in the simulation. The uncertainty due to the QCD calculation, also referred to as the ``scale'' uncertainty, covers the variations of predictions from changing the QCD renormalisation and factorisation scales. The QCD scales are varied independently by factors ranging from one half to two, which in total yields seven different scale choices including the nominal one. The UEPS uncertainty is taken as the difference in the predictions between the \progname{Herwig++}
and the default showering programs.
The fractional theoretical uncertainty in \afactor~for the $qqZZ$ process is about 1.8\%, while the overall uncertainties in the cross-section predictions in the total and fiducial phase spaces are about 3\% and 5\%, respectively. The \progname{Sherpa} generator is used to cross-check the nominal predictions, and the \afactor~factors from \progname{Powheg} and \progname{Sherpa} are consistent with each other within the uncertainty. The \cfactor~predictions from the two generators are found to be consistent within the statistical uncertainty of 1\%, and in this measurement, the theoretical uncertainty in \cfactor~is neglected.
 
The understanding of the \ptll~spectrum in the fiducial phase space is crucial for the study of aTGCs, and the predictions from the two generators differ by up to 10\% for \ptll~around 300 \GeV, which is slightly above the theoretical uncertainty of the \progname{Powheg} prediction. The \progname{Powheg} prediction with the $K$-factors applied has better precision in terms of the EW calculation,
while the \progname{Sherpa} generator is expected to give a better description of $ZZ$ production with extra QCD radiation.
Finally, an uncertainty is applied to the \ptll~prediction, as the sum in quadrature of the theoretical uncertainty estimated with \progname{Powheg} and the difference between \progname{Powheg} and \progname{Sherpa},
which is about 5\% for \ptll{} around 150 \GeV{} and increases to about 11\% for \ptll{} above 250 \GeV.
 
A constant 30\% uncertainty is assigned to the total $ggZZ$ cross-section prediction, which covers the uncertainties concerning the NLO $K$-factor~\cite{Caola:2015psa} and the potential missing higher-order contributions~\cite{Li:2015jva}.
The \afactor~predictions for the $ggZZ$ process from the \progname{gg2vv} and \progname{Sherpa} generators are found to be consistent,
and the \afactor~uncertainty is estimated with \progname{Sherpa} and found to be 4.6\%.
The theoretical uncertainty in \cfactor~is neglected for the $ggZZ$ process.
 
The major experimental uncertainties originate from the luminosity uncertainty, the momentum scale and resolution of leptons and jets, and the lepton reconstruction and selection efficiencies~\cite{Aaboud:2019ynx,Aad:2016jkr,Aaboud:2017jcu,Aaboud:2018ugz}.
Smaller experimental uncertainties are also considered, which include uncertainties due to the trigger selection efficiency,
the $b$-jet identification efficiency, the calculation of the \met~soft-term,
and the variation of the average number of interactions per bunch crossing (hereafter referred to as pile-up uncertainty).
The pile-up uncertainty covers the uncertainty on the ratio between the predicted and measured inelastic cross-section
in the fiducial volume defined by $M_X > 13$ \GeV{} where $M_X$ is the mass of the hadronic system~\cite{Aaboud:2016mmw}.
Overall, the total experimental uncertainty on \cfactor~is 3.1\%, dominated by the jet and lepton components.
The uncertainty in the combined 2015+2016 integrated luminosity is 2.1\%. It is derived, following a methodology similar to that detailed in
\Ref{\cite{DAPR-2013-01}}, and using the LUCID-2 detector for the baseline luminosity measurements~\cite{LUCID2},
from calibration of the luminosity scale using x--y beam-separation scans.

The fractional uncertainties in \afactor~and \cfactor~are summarised in \Tab{\ref{tab:acunc}}. In this analysis, the theoretical uncertainties are treated as uncorrelated between the $qqZZ$ and $ggZZ$ processes, while the experimental uncertainties are considered as fully correlated across the relevant processes and final-state channels.

 
\section{Integrated cross-section results}
\label{sec:obs_intxs}
 
\Tab{\ref{tab:result}} lists separately for the $ee$ and $\mu\mu$ channels the observed data yields and the expectations for
the signal and background contributions after the final selection. \Fig{\ref{fig:met_all}} shows for the combined $ee$ and $\mu\mu$
channels the observed and expected \met{} distributions, which are in good agreement.

\begin{table}[t]
\begin{center}
\renewcommand{\arraystretch}{0.25}
\begin{tabular}[b]{l c c | l c c }
\hline
\hline
\multicolumn{3}{c|}{ \multirow{5}{*}{\afactor}} & \multicolumn{3}{c}{ \multirow{5}{*}{\cfactor}} \\
& & & & & \\
& & & & & \\
& & & & & \\
& & & & & \\
\hline
& \multirow{5}{*}{$qqZZ$}    & \multirow{5}{*}{$ggZZ$}
& & \multirow{5}{*}{$ee$} & \multirow{5}{*}{$\mu\mu$} \\
& & & & & \\
& & & & & \\
& & & & & \\
& & & & & \\
\multirow{5}{*}{Stat.}  & \multirow{5}{*}{1.0\%}  &  \multirow{5}{*}{1.1\%}
& \multirow{4}{*}{Stat.} & \multirow{4}{*}{0.6\%}  & \multirow{4}{*}{0.6\%}   \\
& & & & & \\
& & & & & \\
& & & & & \\
& & & \multirow{4}{*}{Electron}    &  \multirow{4}{*}{2.0\%}    & \multirow{4}{*}{-}   \\
\multirow{5}{*}{PDF}    & \multirow{5}{*}{0.8\%}  & \multirow{5}{*}{3.5\%}
& & & \\
& & & & & \\
& & & & & \\
& & & \multirow{4}{*}{Muon}        &  \multirow{4}{*}{-}        & \multirow{4}{*}{1.9\%} \\
& & & & & \\
\multirow{5}{*}{Scale}  & \multirow{5}{*}{1.4\%}  & \multirow{5}{*}{2.0\%}
& & & \\
& & & & & \\
& & & \multirow{4}{*}{Jet}         & \multirow{4}{*}{2.0\%}     & \multirow{4}{*}{2.0\%} \\
& & & & & \\
& & & & & \\
\multirow{5}{*}{UEPS}   & \multirow{5}{*}{0.1\%}  & \multirow{5}{*}{2.0\%}
& & & \\
& & & \multirow{4}{*}{Soft}        & \multirow{4}{*}{0.9\%}     & \multirow{4}{*}{1.1\%} \\
& & & & & \\
& & & & & \\
& & & & & \\
\hline
\renewcommand{\arraystretch}{0.5}
\multirow{5}{*}{Total} & \multirow{5}{*}{1.9\%} & \multirow{5}{*}{4.6\%}
& \multirow{5}{*}{Total}   & \multirow{5}{*}{3.1\%}  & \multirow{5}{*}{3.1\%}    \\
& & & & & \\
& & & & & \\
& & & & & \\
\hline
\hline
\end{tabular}
\caption{Fractional uncertainties for \afactor{} and \cfactor, with the contributions from
the various sources, theoretical only for \afactor{} and experimental only for
\cfactor.
The uncertainties in \afactor{} for the $qqZZ$ and $ggZZ$ sub-processes are given in different columns.
The uncertainties in \cfactor{} for the $ee$ and $\mu\mu$ channels of
the inclusive $ZZ$ process are given in separate columns.
The total uncertainties in \afactor{} and \cfactor{} are given in the last rows, respectively.
The ``Soft'' term includes the \met~soft-term and the pile-up uncertainties.
}
\label{tab:acunc}
\end{center}
\end{table}

\begin{table}[t]
\begin{center}
\renewcommand{\arraystretch}{1.2}
\begin{tabular}{l | r @{ $\pm$ } r @{ $\pm$ } r | r @{ $\pm$ } r @{ $\pm$ } r}
\hline
\hline
& \multicolumn{3}{c|}{$ee$} & \multicolumn{3}{c}{$\mu\mu$} \\
\hline
Data                & \multicolumn{3}{c|}{371} & \multicolumn{3}{c}{416} \\
\hline
Signal              & \multicolumn{3}{c|}{} & \multicolumn{3}{c}{} \\
$qqZZ$              &  \numRF{194.4}{3} & \numRF{2.9}{1} & \numRF{11.6}{2} & \numRF{202.2}{3} & \numRF{2.8}{1} & \numRF{12.1}{2} \\
$ggZZ$              &   25.1 & 0.3 & 7.7 &  26.4 & 0.3 & 8.1 \\
\hline
Backgrounds         & \multicolumn{3}{c|}{} & \multicolumn{3}{c}{} \\
$WZ$                & 92.9 & 3.0 & 4.8 & 100.7 & 3.2 & 5.2 \\
\Nonresll           & 25.5 & 3.4 & 1.8 &  31.5 & 4.2 & 2.2 \\
\zjet               &  4.7 & 0.2 & 2.3 &   5.9 & 0.3 & 2.8 \\
\ZZllll             &  3.8 & 0.2 & 0.3 &   4.2 & 0.2 & 0.3 \\
Others              & 0.87 & 0.03& 0.17 & 0.87 & 0.03& 0.17   \\
\hline
Background expected  & \numRF{127.8}{3} & \numRF{4.5}{1} & \numRF{5.7}{1} & \numRF{143.2}{3} & \numRF{5.3}{1} & \numRF{6.3}{1} \\
\hline
Total expected      & \numRF{347.3}{3} & \numRF{5.4}{1} & \numRF{15.0}{2} & \numRF{371.8}{3} & \numRF{6.0}{1} & \numRF{15.8}{2} \\
\hline
\hline
\end{tabular}
\renewcommand{\arraystretch}{1}
\caption{Observed data yields and expected signal and background contributions,
shown separately for the $ee$ and $\mu\mu$ channels. The errors shown for the expected yields
correspond to the statistical and systematic contributions in that order.
The expected background and signal+background yields are shown in the last two rows,
where the uncertainties are computed as the sum in quadrature of those from the individual processes.}
\label{tab:result}
\end{center}
\end{table}

\begin{figure}[!htbp]
\begin{center}
\includegraphics[width=0.68\textwidth]{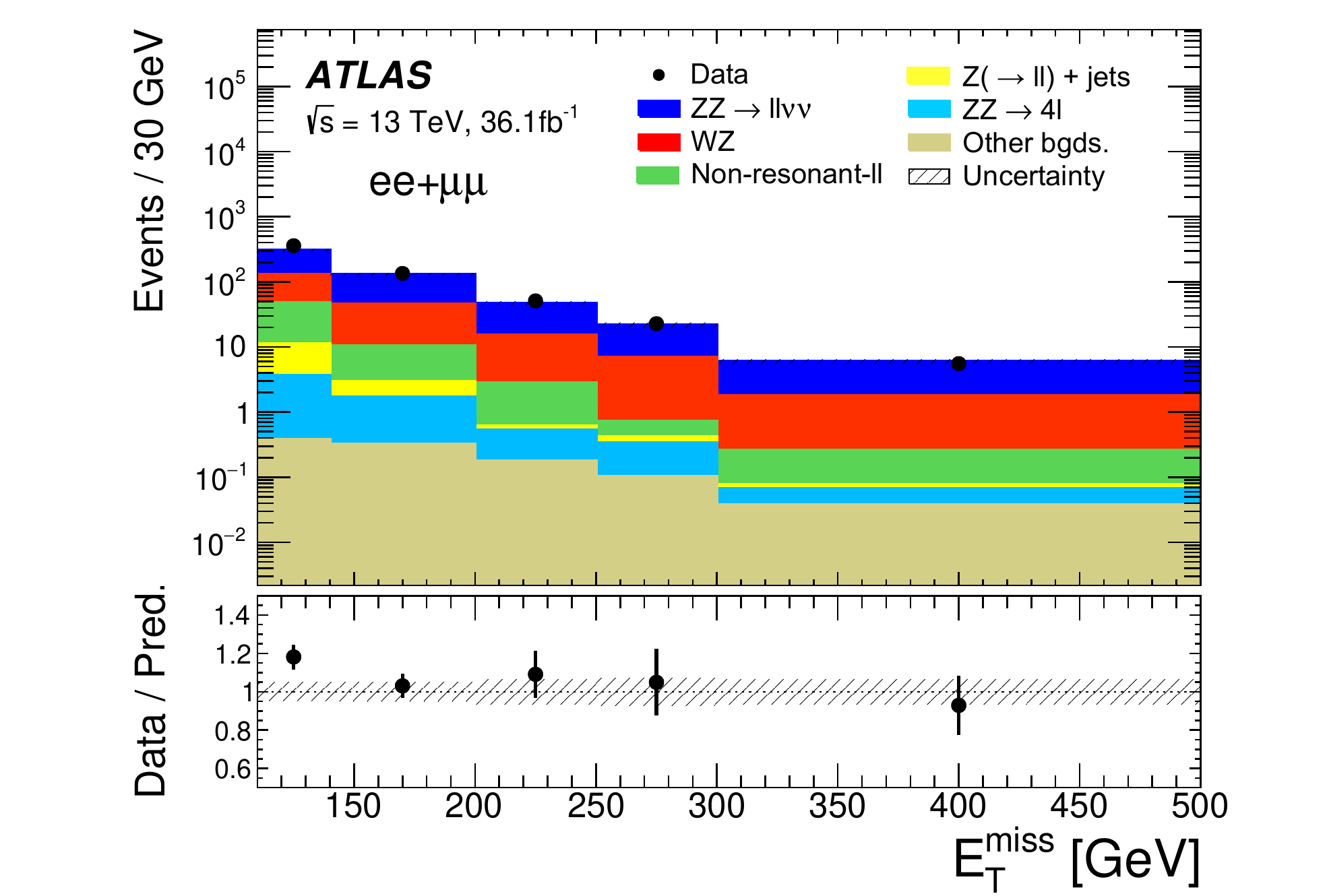}
\end{center}
\caption{Observed and expected \met{} distributions
after the final selection for the combined $ee$ + $\mu\mu$ channel before the fit procedure.
The error bars on the data points correspond to the data statistical uncertainties, and
the hashed band for the prediction includes both the statistical uncertainties of the simulation
and the systematic uncertainties.
The ``other'' background corresponds to the $VVV$ and $t\bar{t}V$ processes.}
\label{fig:met_all}
\end{figure}


The integrated fiducial and total cross-sections (\xsfid{} and \xstot) are determined by binned maximum-likelihood fits to the \met~distributions.
As shown in \Fig{\ref{fig:met_all}}, the signal-to-background ratio increases as \met{} becomes larger.
The use of \met{} improves the precision of the measured fiducial cross-section relatively by 5\% compared
with the case where no kinematic information is used.

The expected yield in each channel $i$ and in each \met{} bin $j$ is given by:
 
\begin{equation*}
N_{\mathrm{exp}}^{ij} = \xsfidEQ \times \mathcal{L} \times C_{ZZ}^i \times f_{ZZ}^{ij} + N_{\mathrm{bkg}}^{ij} \\
= \sigma_{ZZ}^{\mathrm{tot}} \times B \times \mathcal{L} \times A_{ZZ}^i \times C_{ZZ}^i
\times f_{ZZ}^{ij} + N_{\mathrm{bkg}}^{ij},
\end{equation*}
 
where $\mathcal{L}$ is the integrated luminosity, $N_{\mathrm{bkg}}^{ij}$ the expected background yield, $B$ the branching fraction for the \ZZllnn~decay ($\ell = e$ or $\mu$), and $f_{ZZ}^{ij}$ is the fraction of signal events in bin $j$ with respect to the total distribution.
The number of events follows a Poisson distribution in each bin, and the systematic uncertainties are treated as Gaussian nuisance parameters,
$\theta_{k}$, in the fit.
For each source of systematic uncertainty, $k$, a single nuisance parameter is used for all the processes and channels where this uncertainty matters. The statistical uncertainty due to the limited size of simulated samples is treated as uncorrelated among bins and channels.
The binned likelihood function is built over all bins as follows:
 
\begin{equation*}
\mathit{L} (\sigma,\vec{\theta}) = \prod_{i} \prod_{j}
\mathrm{Pois}(N_{\mathrm{obs}}^{ij} | N_{\mathrm{exp}}^{ij}(\sigma,\vec{\theta}))
\times \prod_{k}\mathrm{Gaus}(\theta_{k}),
\end{equation*}
 
where $N_{\mathrm{obs}}^{ij}$ is the observed data yield in each bin.

\Tab{\ref{tab:xsec_unc}} summarises the main sources of uncertainty in the measured combined fiducial cross-section,
where individual sources of a similar nature are grouped together. 
The statistical and total systematic uncertainties in the measurement are of similar sizes.
\Tab{\ref{tab:results_xsec}} shows the measured fiducial cross-sections, separately for each channel and for their combination,
together with the breakdown of their uncertainties. The $ee$ and $\mu\mu$ channel cross-sections are compatible within their
respective statistical uncertainties. The measured combined fiducial cross-section has a total uncertainty of 7\%,
which is significantly better than the previous measurement~\cite{STDM-2014-16},
and comparable in size to that obtained in the $ZZ\to 4\ell$ channel~\cite{STDM-2016-15,CMS-SMP-16-017}.
\Tab{\ref{tab:results_xsec}} also shows the combined measured total cross-section, as well as the predictions for
the cross-sections, as taken from \Tab{\ref{tab:psnum}}.
The combined measurement is about 13\% higher than the prediction, which is not significant given the size of
the measurement and prediction uncertainties.

\begin{table}[t]
\begin{center}
\renewcommand\arraystretch{1.2}
\begin{tabular}{c |c c| c c c c}
\hline
\hline
& & &   Lumi. & Electron & Muon  & Jet    \\
Total & Data stat. & Total syst. &    2.2\% & 1.2\%    & 1.1\% & 2.1\%  \\\cline{4-7} 
7.0\% & 5.5\% & 4.3\% &   $WZ$  & Non-resonant-$\ell\ell$ & \zjet & Sim. stat.\\
& & &   1.6\% & 1.6\%     & 0.4\% & 0.7\% \\
\hline
\hline
\end{tabular}
\renewcommand\arraystretch{1}
\caption{Relative contributions to the measured combined fiducial cross-section from the main sources of uncertainty
after the fit procedure.
The total uncertainty includes the data statistical and systematic components.
For the systematic uncertainty, the individual sources of a similar nature are grouped together for simplicity.
``Sim. stat.'' indicates the uncertainty source corresponding to the limited size of the simulation samples for the signal
and background processes.
}
\label{tab:xsec_unc}
\end{center}
\end{table}

\begin{table}[t]
\begin{center}
\renewcommand\arraystretch{1.4}
\begin{tabular}{l c c c}
\hline
\hline
& & Measured & Predicted \\
\hline
\multirow{3}{*}{\xsfid~[fb]} &  $ee$ & 12.2 $\pm$ 1.0 (stat) $\pm$ 0.5 (syst) $\pm$ 0.3 (lumi) & 11.2 $\pm$ 0.6 \\
&  $\mu\mu$    & 13.3 $\pm$ 1.0 (stat) $\pm$ 0.5 (syst) $\pm$ 0.3 (lumi) & 11.2 $\pm$ 0.6 \\
&  $ee+\mu\mu$  & 25.4 $\pm$ 1.4 (stat) $\pm$ 0.9 (syst) $\pm$ 0.5 (lumi) & 22.4 $\pm$ 1.3 \\
\hline
\xstot~[pb] & Total & 17.8 $\pm$ 1.0 (stat) $\pm$ 0.7 (syst) $\pm$ 0.4 (lumi) & 15.7 $\pm$ 0.7 \\
\hline
\hline
\end{tabular}
\renewcommand\arraystretch{1}
\caption{Measured and predicted integrated cross-sections in the fiducial and total phase spaces,
together with the breakdown of their uncertainties. The luminosity uncertainty is quoted separately
from the other systematic uncertainties. The measurements are also shown separately
for the $ee$ and $\mu\mu$ channels in the case of the fiducial cross-section.}
\label{tab:results_xsec}
\end{center}
\end{table}

 
\section{Differential cross-section results}
\label{sec:diffxs}
 
Differential cross-sections are measured in the fiducial phase space by counting data events observed in each bin of the observables of interest,
after subtracting the expected background contribution, and correcting for the detector effects with the unfolding procedure,
chosen here to be the iterative Bayesian unfolding method of \Ref{\cite{DAgostini:1994fjx}}.
The unfolding process takes into account fiducial corrections (correcting for events outside the fiducial phase space but passing
the detector-level selections), bin-to-bin migrations due to detector resolution, and detector inefficiencies.
An optimal number of two iterations is used for this analysis, as a balance between the size of the statistical uncertainty
in the measurement and residual biases from the method. The residual bias is in almost all bins below 1\%,
as estimated by comparing the results obtained using different prior distributions (constant, expected, observed)
in the unfolding process.

The experimental uncertainties for the measurement results are evaluated by varying the response matrices for the unfolding
according to the $\pm 1\sigma$ effects of each uncertainty source, and by comparing the resulting unfolded results
with the nominal one.
The background uncertainties are considered at the stage of the background subtraction. The statistical uncertainty of the data is estimated by repeating the unfolding procedure with 2000 sets of pseudo-data and then taking the root mean square of the deviations of the resulting spectra from the data spectrum.
The response matrices are also subject to the statistical uncertainty of the simulated samples, which is estimated using a similar approach.
 
Figures~\ref{fig:unfold_results_1} and \ref{fig:unfold_results_2} present the measured differential cross-sections for the eight
observables of interest defined in \Sect{\ref{sec:intro}}. The binning for each variable is chosen to minimise the bin-to-bin migrations
while preserving a sufficient number of events per bin.
The \njets{} spectrum is measured only for hard jets as defined in \Sect{\ref{sec:sel}},
and the \ptleadj{} distribution is obtained in the fiducial phase space of events containing at least one hard jet.
The predictions from \progname{Powheg+gg2vv} and \progname{Sherpa} are also shown in Figures~\ref{fig:unfold_results_1} and
\ref{fig:unfold_results_2}, and are found to be in agreement with the measurements within uncertainties.
The electroweak production of $ZZ$ associated with two jets is not taken into account in the predictions due to its negligible contribution.
The differential measurements are largely dominated by the statistical uncertainty on the data, but the systematic uncertainties contribute significantly in certain regions of phase space.
As an example, the uncertainties from the various sources for the differential measurement of the \ptll{} distribution are listed
in Table~\ref{tab:unfold_ptll_uncer_break}.
 
Comparing with the Run-1 results of \Ref{\cite{STDM-2014-16}},
this measurement is obtained from a larger dataset with highly improved accuracy and for a wider range of observables.

\begin{figure}[t]
\begin{center}
\includegraphics[width=0.495\textwidth]{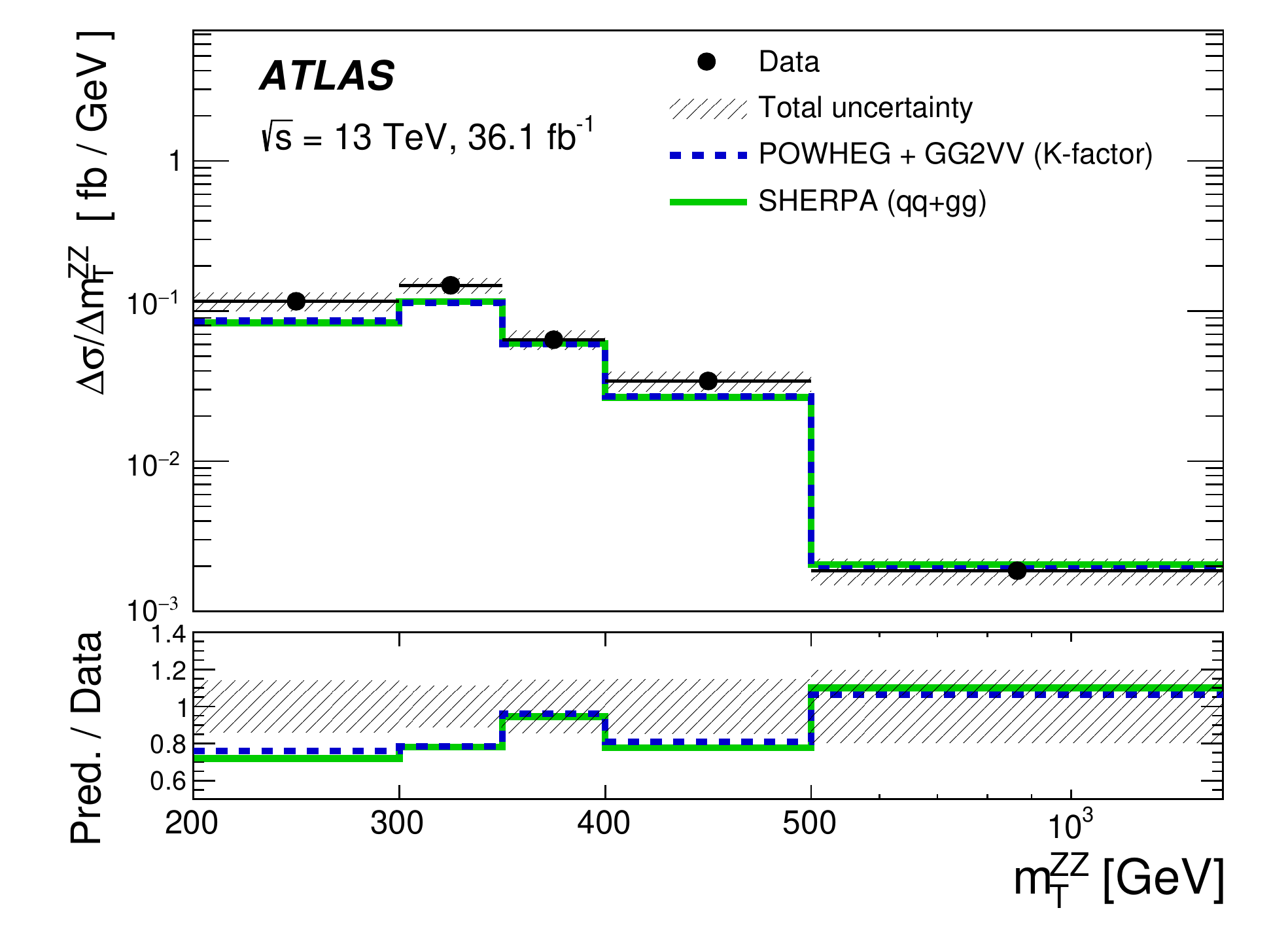}
\includegraphics[width=0.495\textwidth]{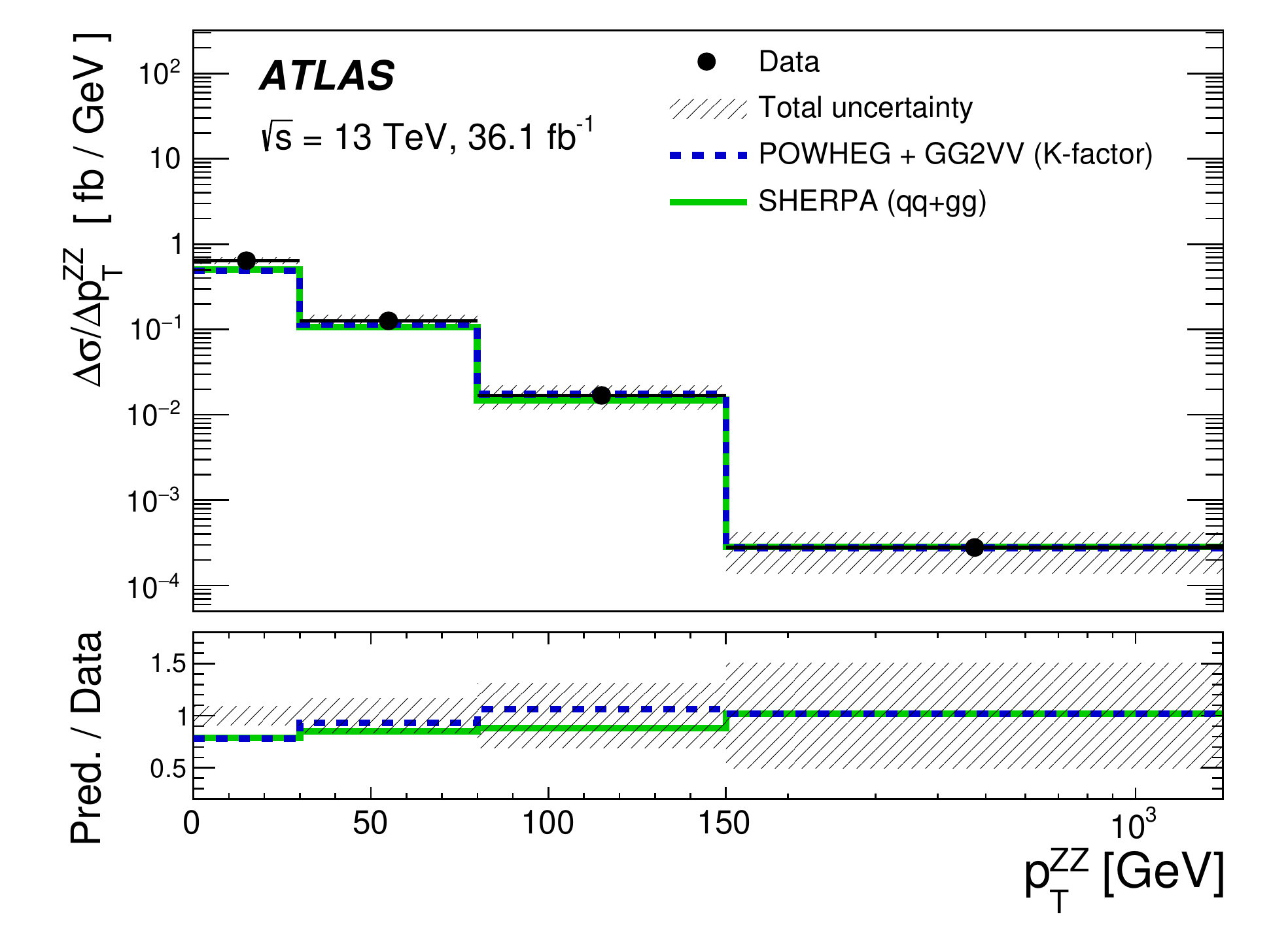}
\includegraphics[width=0.495\textwidth]{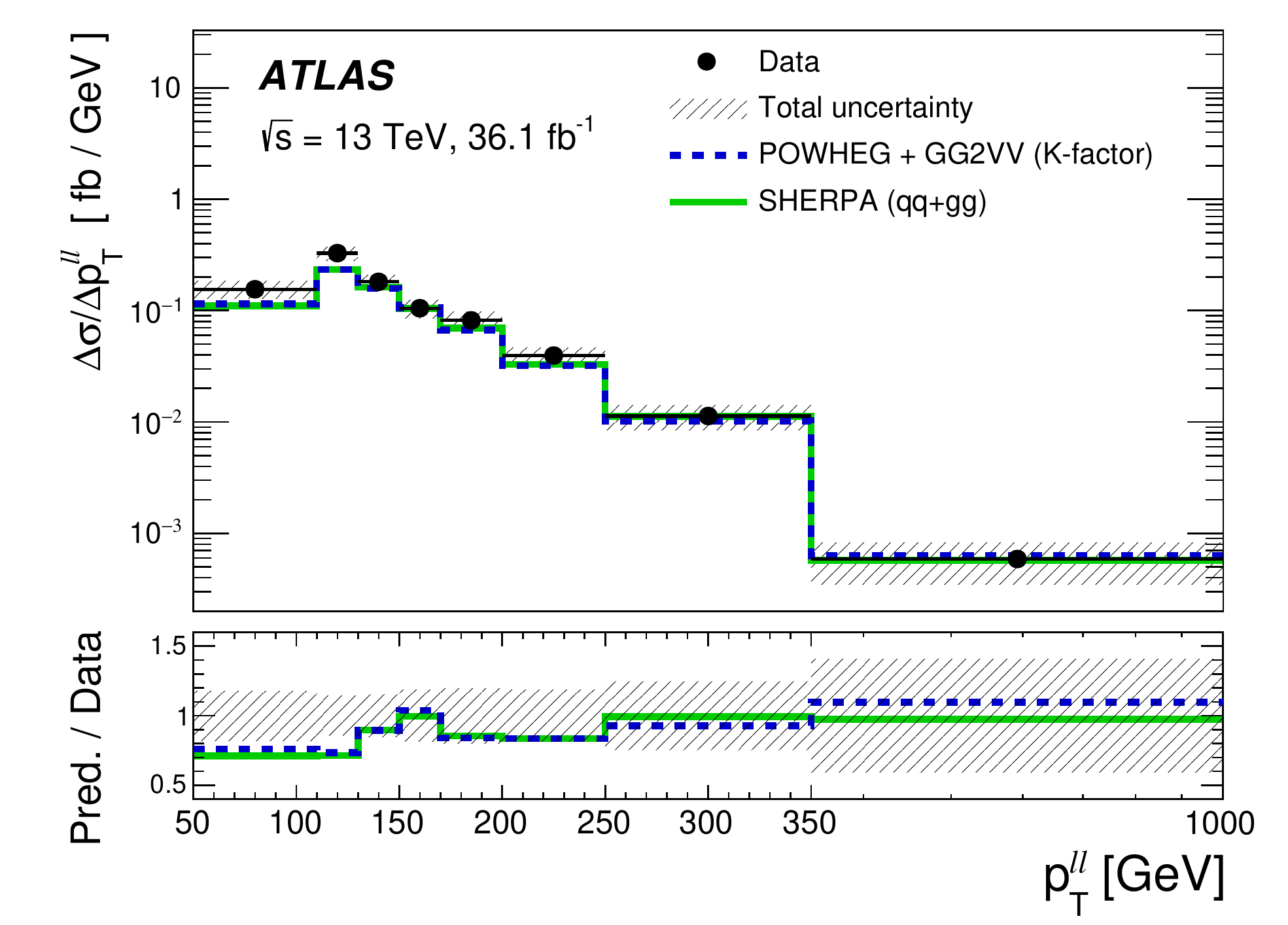}
\includegraphics[width=0.495\textwidth]{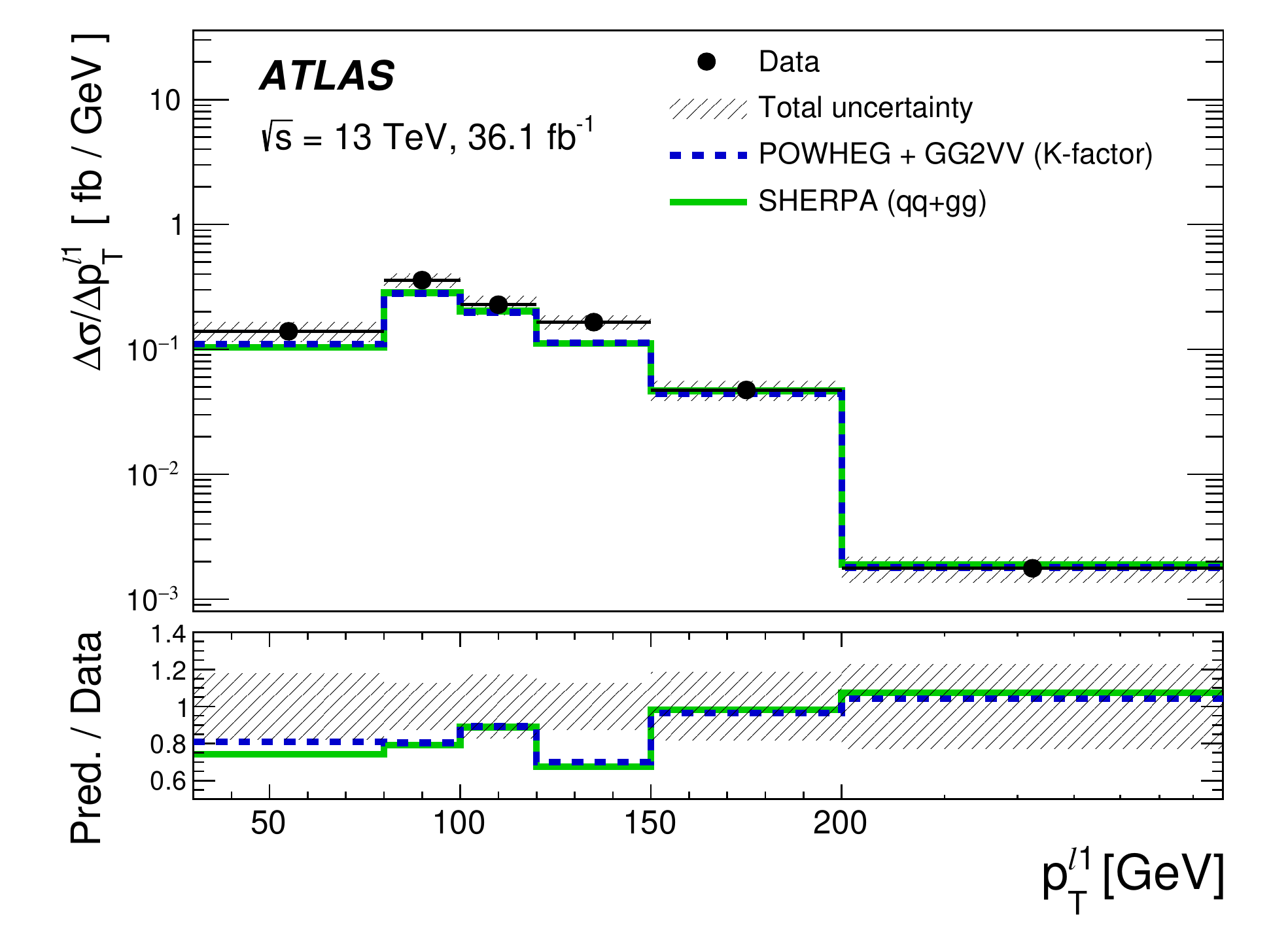}
\end{center}
\caption{The measured differential cross-sections for \mtZZ{} (top left), \ptZZ{} (top right), \ptll{} (bottom left),
and \ptleadl{} (bottom right) in the fiducial region.
The hashed bands in light grey give the total uncertainty.
Also shown are the predictions as obtained from \progname{Sherpa} (solid line) and \progname{Powheg+gg2vv}
(dashed line) with higher-order $K$-factors, as described in \Sect{\ref{sec:datamc}}.
The last bin in the four distributions is shown using a different $x$-axis scale for better visualisation.
}
\label{fig:unfold_results_1}
\end{figure}

\begin{figure}[t]
\begin{center}
\includegraphics[width=0.495\textwidth]{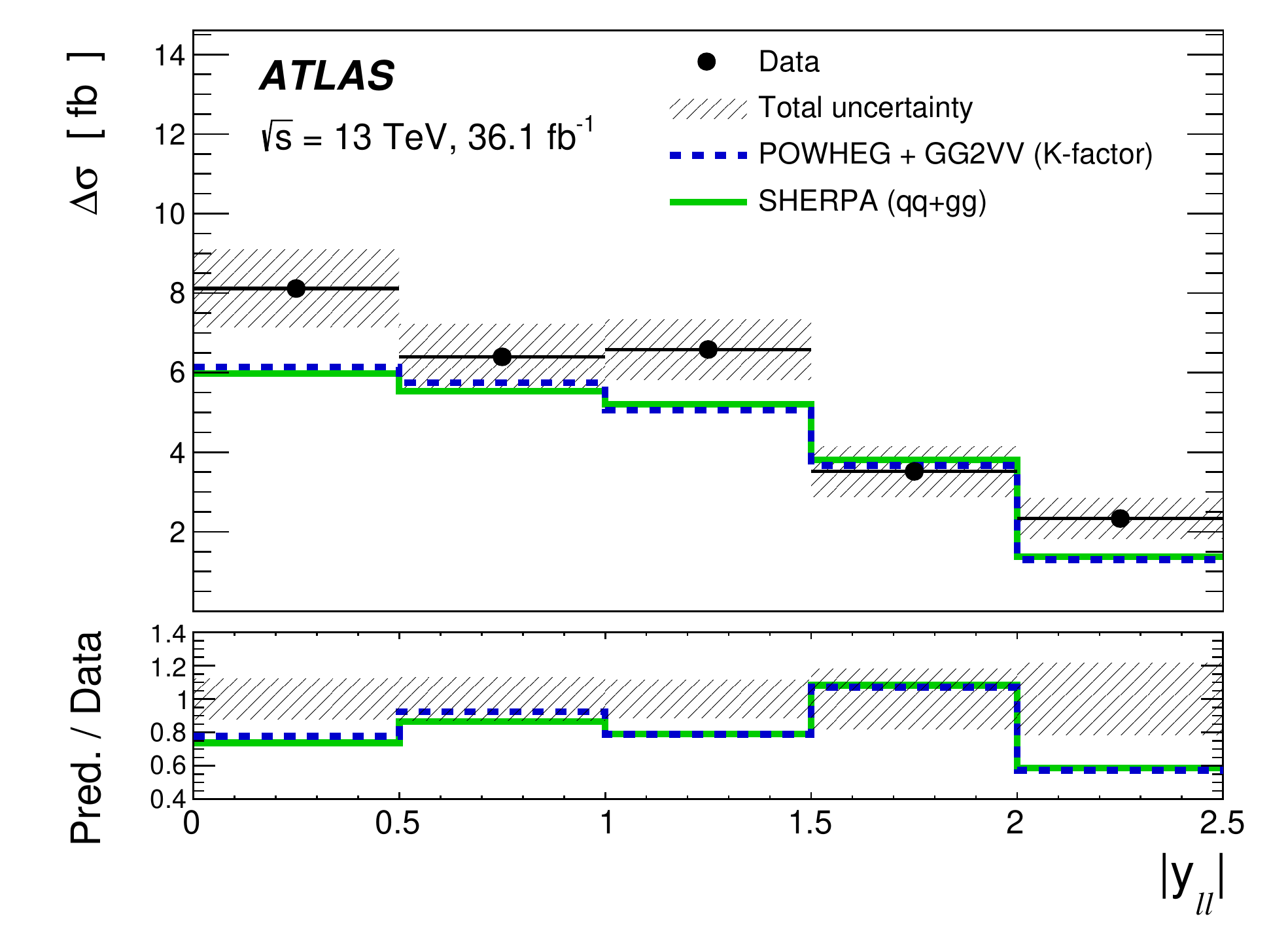}
\includegraphics[width=0.495\textwidth]{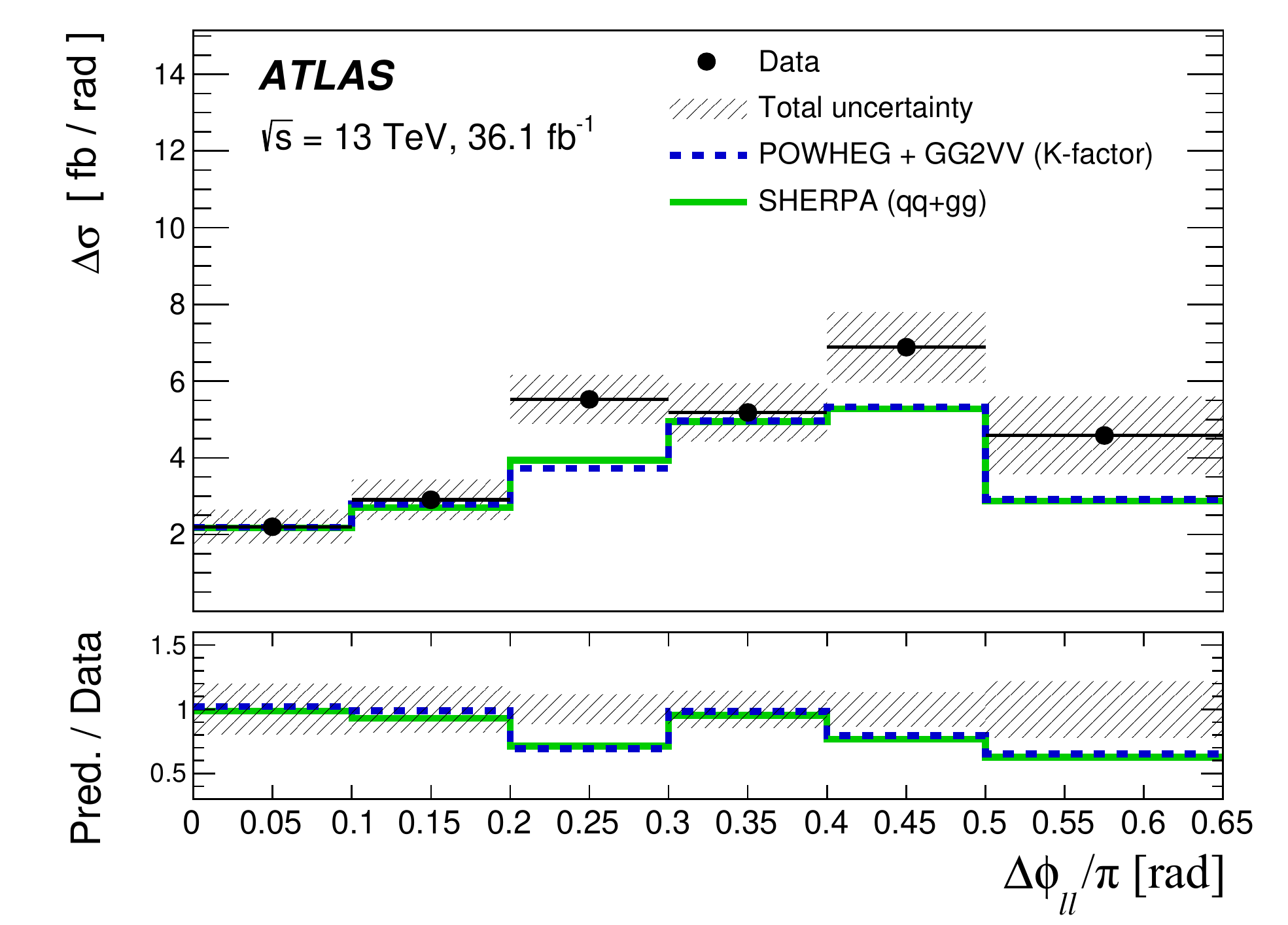}
\includegraphics[width=0.495\textwidth]{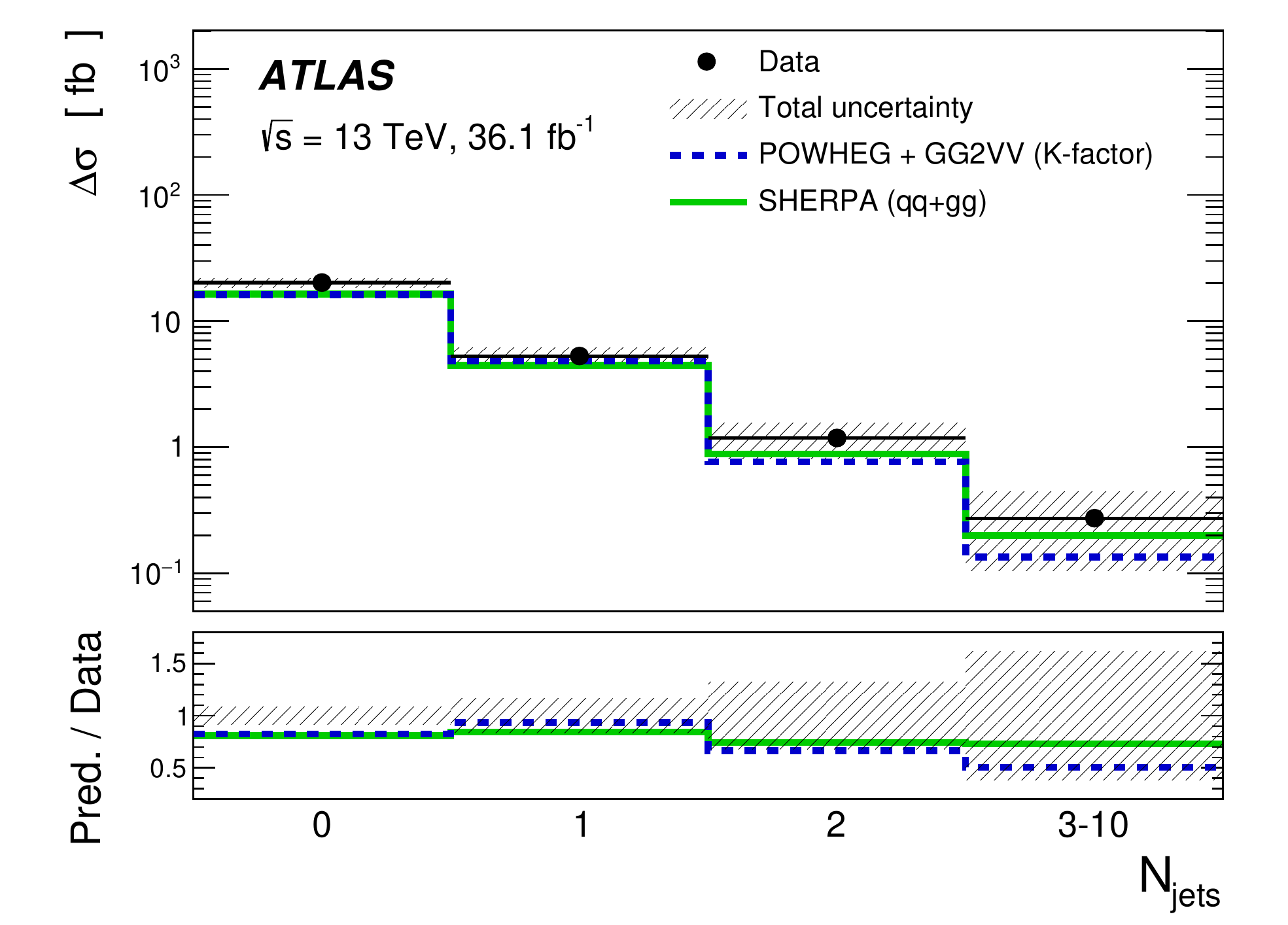}
\includegraphics[width=0.495\textwidth]{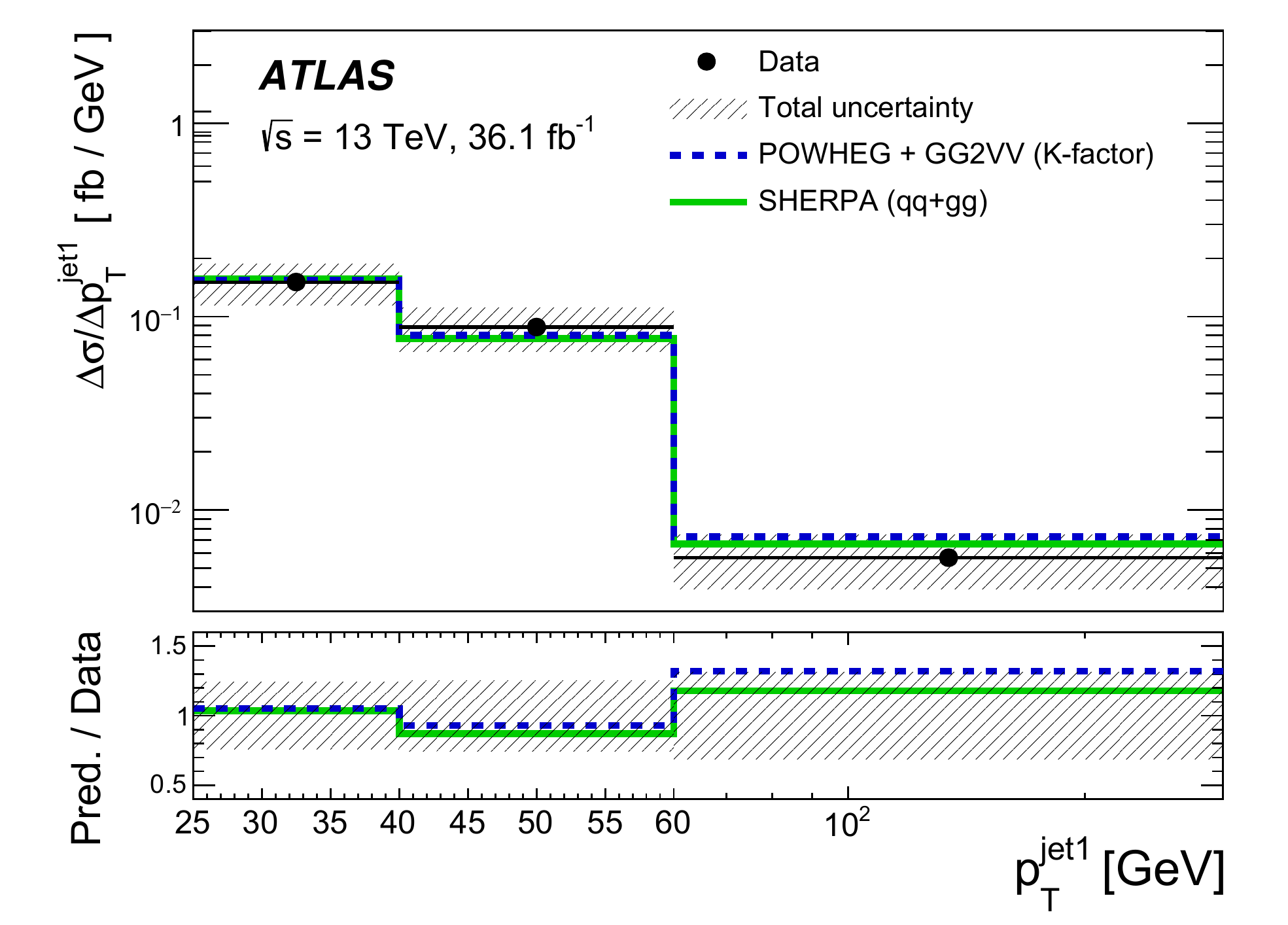}
\end{center}
\caption{The measured differential cross-sections for \yllabs{} (top left), \dphill{} (top right), \njets{} (bottom left),
and \ptleadj{} (bottom right) in the fiducial region.
The hashed bands in light grey give the total uncertainty.
Also shown are the predictions as obtained from \progname{Sherpa} (solid line) and \progname{Powheg+gg2vv}
(dashed line) with higher-order $K$-factors, as described in \Sect{\ref{sec:datamc}}.
The last bin in the distributions of \njets{} and \ptleadj{}
is shown using a different $x$-axis scale for better visualisation.
}
\label{fig:unfold_results_2}
\end{figure}

\begin{table}[!htbp]
\center
\resizebox*{0.9\textwidth}{!}{
\begin{tabular}{l | r@{.}l r@{.}l r@{.}l r@{.}l r@{.}l r@{.}l r@{.}l r@{.}l}
\hline\hline
\ptll{} range [\GeV] &  \multicolumn{2}{c}{50--110}  & \multicolumn{2}{c}{110--130} & \multicolumn{2}{c}{130--150} &
\multicolumn{2}{c}{150--170} & \multicolumn{2}{c}{170--200} & \multicolumn{2}{c}{200--250} &
\multicolumn{2}{c}{250--350} & \multicolumn{2}{c}{350--1000}   \\
\hline
Measured $\sigma$ (fb) & 9 & 3   & 6 & 6  & 3 & 6  & 2 & 1  & 2 & 5  & 2 & 0  & 1 & 1  & 0 & 4   \\
Total unc. & 17 & 7 \% &  13 & 6 \%  & 15 & 2 \% & 18 & 6 \%  & 18 & 6 \% & 17 & 6 \% & 24 & 9 \%  & 40 & 5 \%  \\
\hline
Stat. unc. & 14 & 7 \%  & 11 & 1 \% & 14 & 0 \% & 17 & 7 \% & 16 & 0 \% & 16 & 9 \% & 23 & 4 \% & 39 & 4 \%  \\
Syst. unc. & 7 & 0  \%  & 4 & 5 \% & 5 & 0 \% & 4 & 3 \% & 3 & 9 \% & 4 & 6 \% & 4 & 6 \% & 5 & 5 \%  \\
Bkg. unc. & 6 & 9 \%  & 6 & 4 \% & 3 & 2 \% & 3 & 7 \% & 8 & 6 \% & 2 & 1 \% & 7 & 1 \% & 7 & 6 \%  \\
\hline
Sim. stat. & 1 & 2 \%  & 0 & 7 \% & 0 & 7 \% & 0 & 8 \% & 0 & 9 \% & 0 & 9 \% & 1 & 1 \% & 2 & 0 \%  \\
Electron & 0 & 7 \%  & 0 & 8 \% & 0 & 9 \% & 1 & 7 \% & 1 & 3 \% & 1 & 6 \% & 2 & 1 \% & 3 & 2 \%  \\
Muon & 1 & 0 \%  & 1 & 3 \% & 1 & 0 \% & 1 & 1 \% & 1 & 2 \% & 1 & 4 \% & 2 & 0 \% & 1 & 7 \%  \\
Jet & 5 & 4 \%  & 2 & 9 \% & 3 & 8 \% & 3 & 0 \% & 2 & 3 \% & 2 & 1 \% & 2 & 7 \% & 2 & 5 \%  \\
Soft & 3 & 6 \%  & 2 & 2 \% & 2 & 0 \% & 0 & 8 \% & 1 & 3 \% & 2 & 7 \% & 0 & 3 \% & 1 & 7 \% \\
Luminosity & 2 & 1 \%  & 2 & 1 \% & 2 & 1 \% & 2 & 1 \% & 2 & 1 \% & 2 & 1 \% & 2 & 1 \% & 2 & 1 \% \\
\hline\hline
\end{tabular}
}
\caption{Measured cross-sections and breakdown of uncertainties (\%) for the unfolded
$p_{\text{T}}^{\ell\ell}$ distribution in the fiducial region.
The top part of the table gives separately the three main contributions to the total uncertainty,
arising respectively from data statistics (labelled Stat.), background subtraction (labelled Bkg.),
and other systematic uncertainties (labelled Syst.).
The bottom part of the table shows a more detailed breakdown of the third contribution (Syst.).
The ``Soft'' term includes the \met~soft-term and the pile-up uncertainties.
}
\label{tab:unfold_ptll_uncer_break}
\end{table}

 
\section{Search for aTGCs}
\label{sec:tgc}
 
The search for aTGCs is carried out using the unfolded \ptll{} distribution of \Fig{\ref{fig:unfold_results_1}} in the fiducial phase space.
The contribution due to aTGCs is introduced using an effective vertex function approach~\cite{Baur:2000ae}.
It includes two coupling parameters that violate charge-parity (CP) symmetry, $f_4^\gamma$ and $f_4^Z$,
as well as two CP-conserving ones, $f_5^\gamma$ and $f_5^Z$.
The sensitivity range of this search is found to be within the unitarity bounds,
so no form factors~\cite{Baur:2000ae} are applied to the coupling parameters in this analysis. Furthermore,
the coupling parameters are used to extract information about the dimension-eight operators of the effective field theory~\cite{Degrande:2013kka}.
 
The relative change in the SM production cross-section in each \ptll{} bin is parameterised as a quadratic function
depending on the coupling parameters.
The parameterisation for the $qqZZ$ process is derived from simulation as described in Section~\ref{sec:datamc},
while the impact of the aTGCs is neglected for the $ggZZ$ process.
Since the sensitivity to possible aTGCs lies in the high-\ptll{} region, only the bins with \ptll~$>150$~\GeV{} are considered in the search.
\Fig{\ref{fig:aTGC_pTZ}} compares the measured \ptll{} spectrum in this region of phase space to the SM prediction alone and
to the SM prediction augmented with aTGCs corresponding to different values of the coupling parameters described above.
The highest-\ptll{} bins are required to contain a minimum of ten observed events at the detector level, which arises from the fact
that the limits on aTGCs derived from the unfolded distribution rely on a Gaussian approximation for the statistical fluctuations from the data.
The total uncertainty in the measured cross-section in the last bin is about 40\%,
while the corresponding SM prediction has an uncertainty of 15\%.
The sensitivity to aTGCs is thus still limited by the statistical uncertainty of the data.
 
Since the observation is consistent with the SM prediction, 95\% confidence intervals (CIs) for the coupling parameters are derived. The test statistic is based on a profile log-likelihood ratio~\cite{cowan:Asymptotic2013},
and the likelihood function is described by a multivariate Gaussian distribution with the following form:
 
\begin{equation*}\label{eq:stat}
\begin{aligned}
L =   \frac{1}{\sqrt{(2\pi)^{5}|\mathcal{C}|}} \times
\mathrm{exp}(-\frac{1}{2} (\vec{x} - \vec{\mu})^{T} \mathcal{C}^{-1} (\vec{x} - \vec{\mu}))
\times \prod_{s}\text{Gaus}(\theta_{s}),
\end{aligned}
\end{equation*}
 
where $\vec{x}$ represents the measured \ptll{} spectrum, $\vec{\mu}$ the expected one,
and $\mathcal{C}$ is the expected covariance matrix
in the differential measurement, with the term $|\mathcal{C}|$ standing for
the determinant of the covariance matrix.
Both $\vec{x}$ and $\vec{\mu}$ have five elements, corresponding to the five \ptll{} bins used in this search.
The expectation $\vec{\mu}$ depends on the coupling parameters and a set of Gaussian nuisance parameters, $\theta_{s}$,
which encapsulate the theoretical uncertainties in the prediction.
The expected covariance matrix $\mathcal{C}$ is first derived by unfolding the predicted SM spectrum at detector level
and then modified to account for the fact that the prediction $\vec{\mu}$, varied in the statistical test,
differs from the SM prediction $\vec{\mu_0}$ in the fiducial region in the case of non-zero coupling parameters.
The relative difference between $\vec{\mu}$ and $\vec{\mu_0}$ can be expressed using a vector $\vec{k}$,
with $k^i = \mu^i/\mu_0^i$ for bin-$i$.
Each matrix element $\mathcal{C}^{ij}$ is the sum of a statistical uncertainty component ($\mathcal{C}^{ij}_{\mathrm{stat}}$)
and systematic uncertainty components arising from SM $ZZ$ process ($\mathcal{C}^{ij}_{ZZ-\mathrm{syst}}$)
and background contribution ($\mathcal{C}^{ij}_{\mathrm{bkg-syst}}$);
the dependence of $\mathcal{C}$ on $\vec{\mu}$ is implemented such that
$\mathcal{C}^{ij}_{\mathrm{stat}}$ and $\mathcal{C}^{ij}_{ZZ-\mathrm{syst}}$ scale with $\sqrt{k^{i}\times k^{j}}$ and $k^{i}\times k^{j}$,
respectively.

The 95\% CIs are derived by scanning the parameter space and using a frequentist method with the CL$_\mathrm{s}$ formalism~\cite{Read:2002hq}.
In the first step, the one-dimensional CI for each aTGC parameter is derived with all other parameters set to zero,
and the results are shown in \Tab{\ref{tab:results_aTGC_1D}}.
These intervals are found to be more stringent than those obtained from the $ZZ\to 4\ell$ channel based on the same
ATLAS data sample~\cite{STDM-2016-15}. In the second step, two-dimensional CIs for each pair of coupling parameters are derived
with the other two parameters set to zero. \Fig{\ref{fig:aTGC_2D_Limits}} presents the two-dimensional CI contours
for the six possible pairs of aTGC coupling parameters.
Finally, in the context of effective field theories, a one-dimensional 95\% CI can be placed on the CP-even dimension-eight
operator $C_{\bar{B}W}/\Lambda^4$~\cite{Degrande:2013kka}.
This is obtained by extrapolating the results of \Tab{\ref{tab:results_aTGC_1D}} following the conversion formula
in \Ref{\cite{Degrande:2013kka}}.
A 95\% CI of [$-$4.0, 4.0] in units of \TeV$^{-4}$ is set for $C_{\bar{B}W}/\Lambda^4$,
which is improved by 30\% with respect to the ATLAS results in the $ZZ\to 4\ell$ channel~\cite{STDM-2016-15}.

\begin{figure}[t]
\begin{center}
\includegraphics[width=0.495\textwidth]{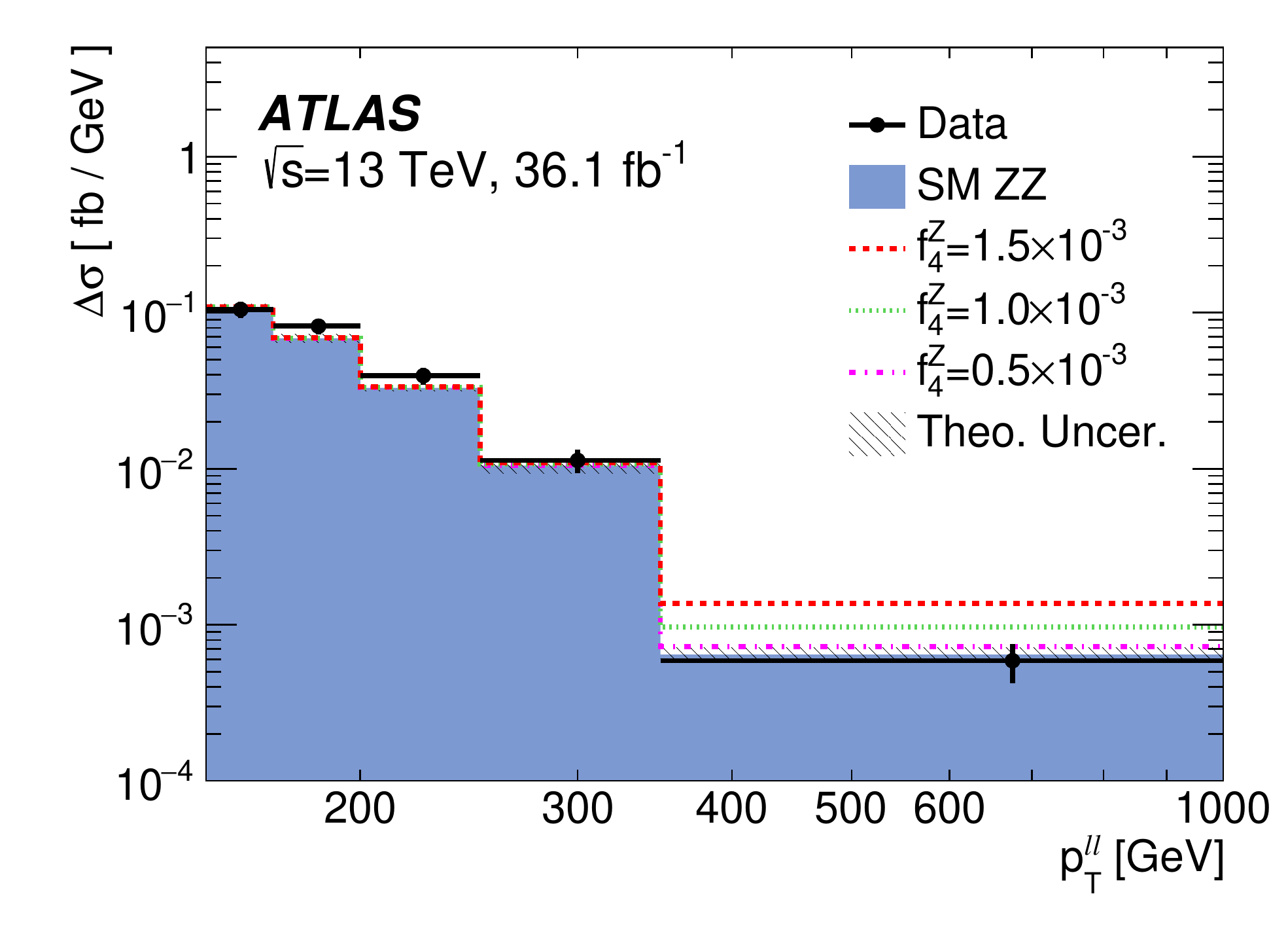}\label{fig:aTGC_f4Z_pTZ}
\includegraphics[width=0.495\textwidth]{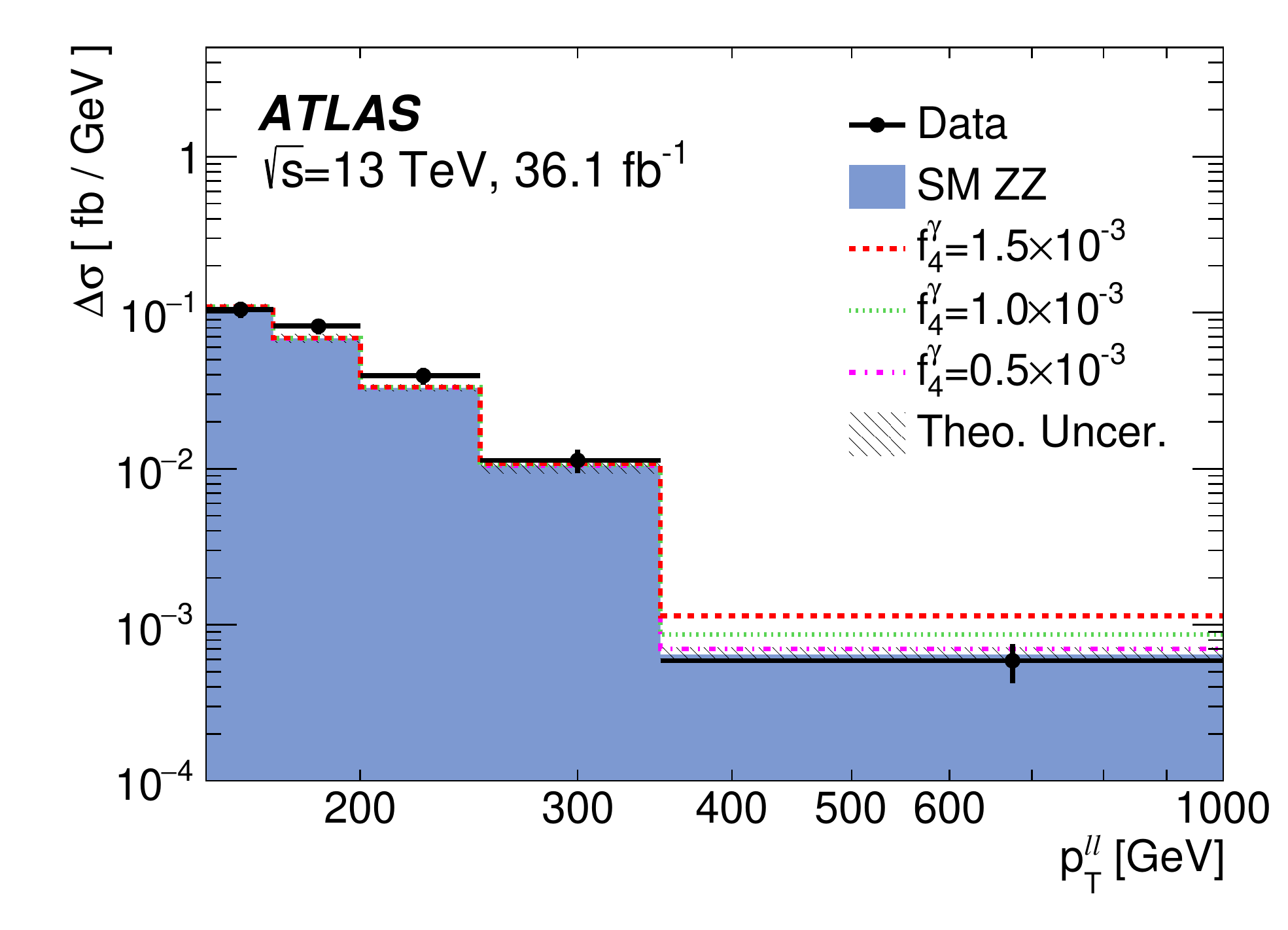}\label{fig:aTGC_f4g_pTZ}
\includegraphics[width=0.495\textwidth]{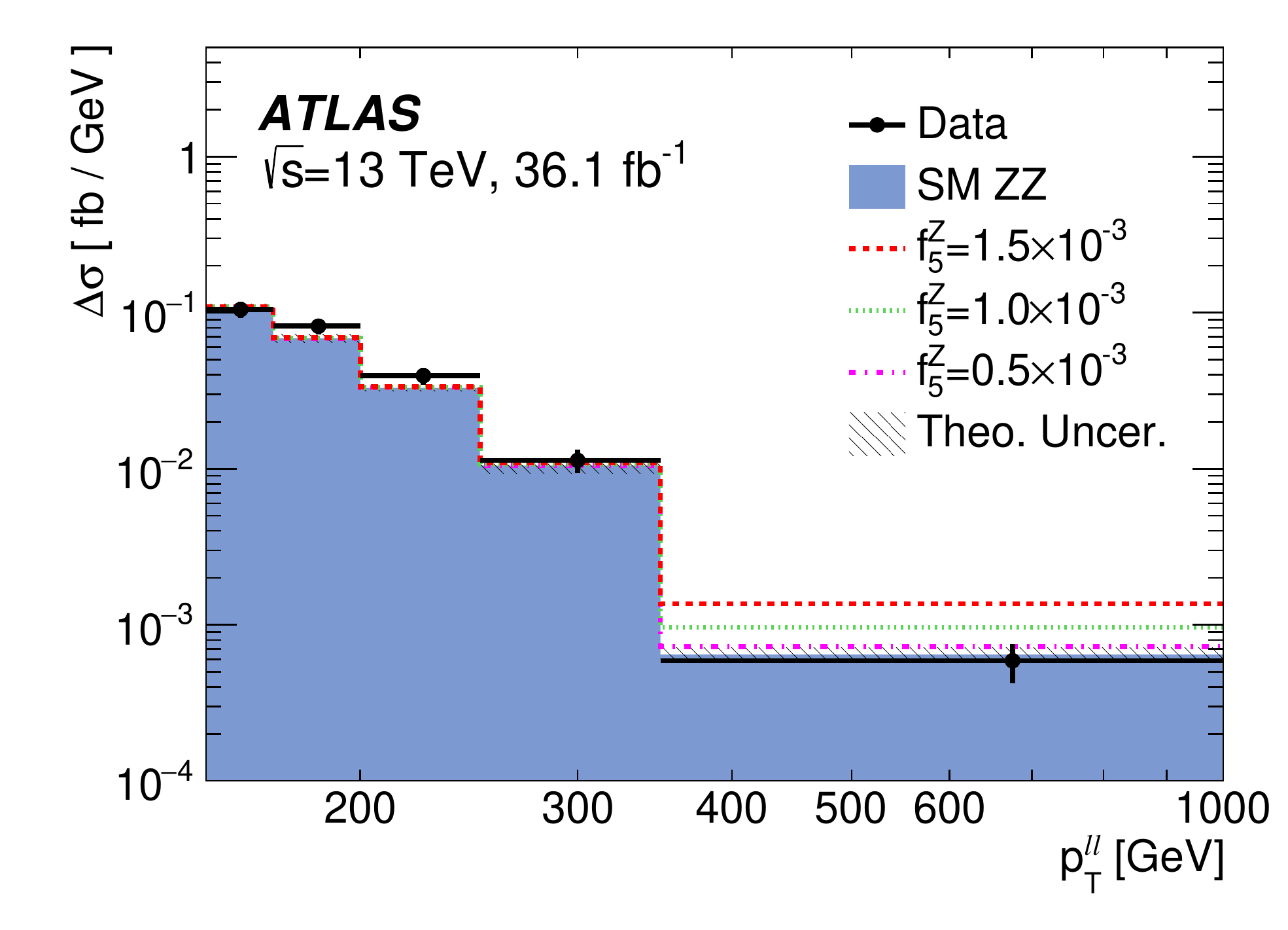}\label{fig:aTGC_f5Z_pTZ}
\includegraphics[width=0.495\textwidth]{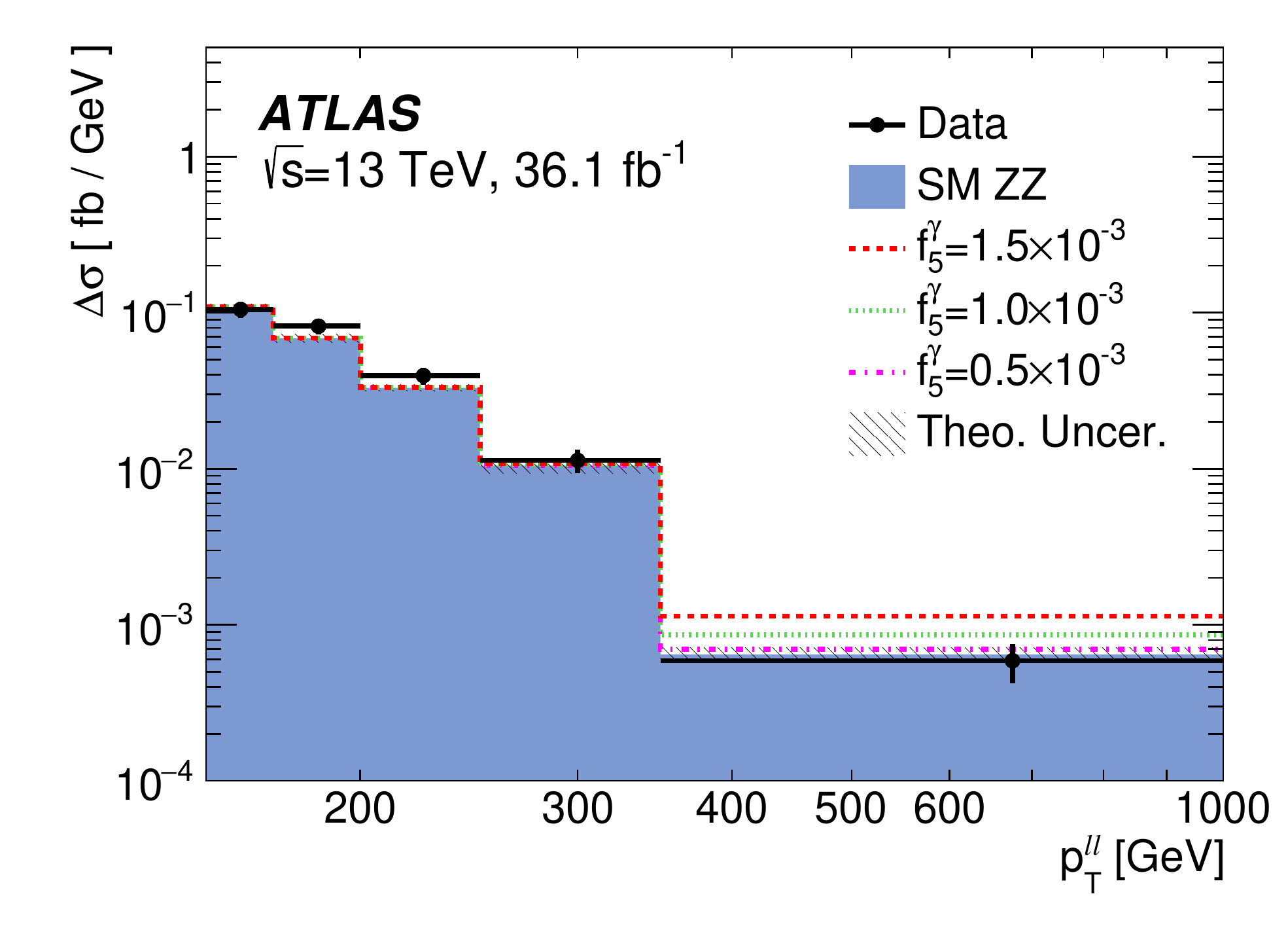}\label{fig:aTGC_f5g_pTZ}
\end{center}
\caption{Unfolded measured distribution of \ptll{} for the bins with \ptll~$>150$~\GeV.
The data are compared with the SM prediction from \progname{Powheg+gg2vv} with higher-order $K$-factors
to which are added possible aTGC contributions for different values of the strength of the coupling parameters
defined in the text.
The results are shown separately for $f_4^{Z}$ (top left), $f_4^{\gamma}$ (top right),
$f_5^{Z}$ (bottom left), and $f_5^{\gamma}$ (bottom right).
The uncertainty band represents the theoretical uncertainties in the SM predictions.
}
\label{fig:aTGC_pTZ}
\end{figure}
 
\begin{table}[t]
\begin{center}
 
\renewcommand\arraystretch{1.8}
\begin{tabular}{l|c|c|c|c}
\hline
\hline
& $f_4^\gamma$ & $f_4^{\text{Z}}$ & $f_5^\gamma$ & ${f_5^\text{Z}}$ \\
\hline
Expected [$\times 10^{-3}$] &  [\numRF{-1.30}{2}, \numRF{1.29}{2}] &[\numRF{-1.11}{2}, \numRF{1.11}{2}]  &[\numRF{-1.30}{2}, \numRF{1.30}{2}] &[\numRF{-1.11}{2}, \numRF{1.11}{2}] \\
Observed [$\times 10^{-3}$]&  [\numRF{-1.21}{2}, \numRF{1.20}{2}] &[\numRF{-1.03}{2}, \numRF{1.03}{2}]  &[\numRF{-1.21}{2}, \numRF{1.20}{2}] &[\numRF{-1.03}{2}, \numRF{1.03}{2}] \\
\hline
\hline
\end{tabular}
\caption{One-dimensional 95\% confidence intervals for the aTGC parameters described in the text.
The limits on each parameter are derived with the other parameters set to zero.}
\label{tab:results_aTGC_1D}
\end{center}
\end{table}

\begin{figure}[!htbp]
\begin{center}
\includegraphics[width=0.495\textwidth]{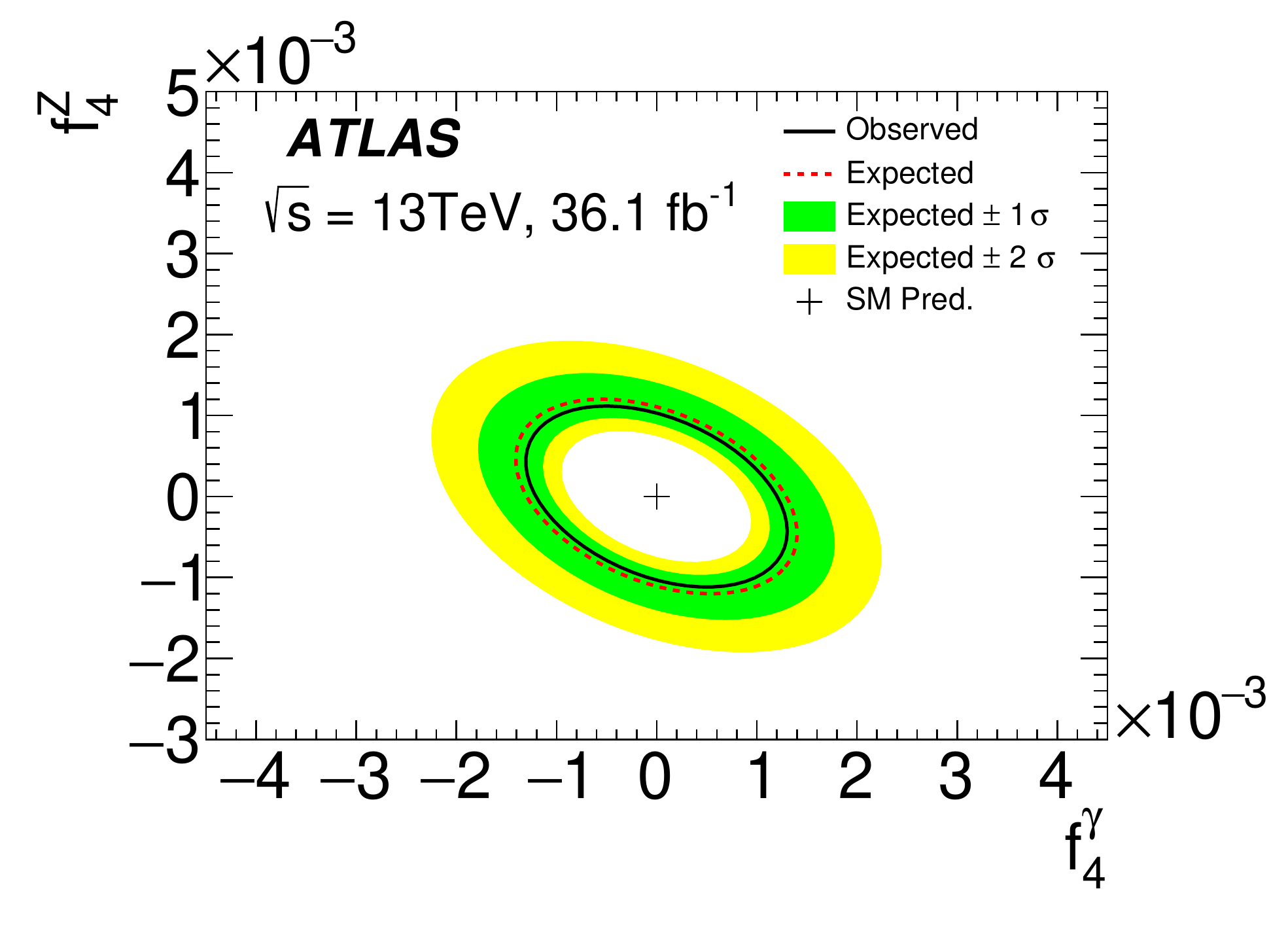}\label{fig:aTGC_f4g_f4Z_2D}
\includegraphics[width=0.495\textwidth]{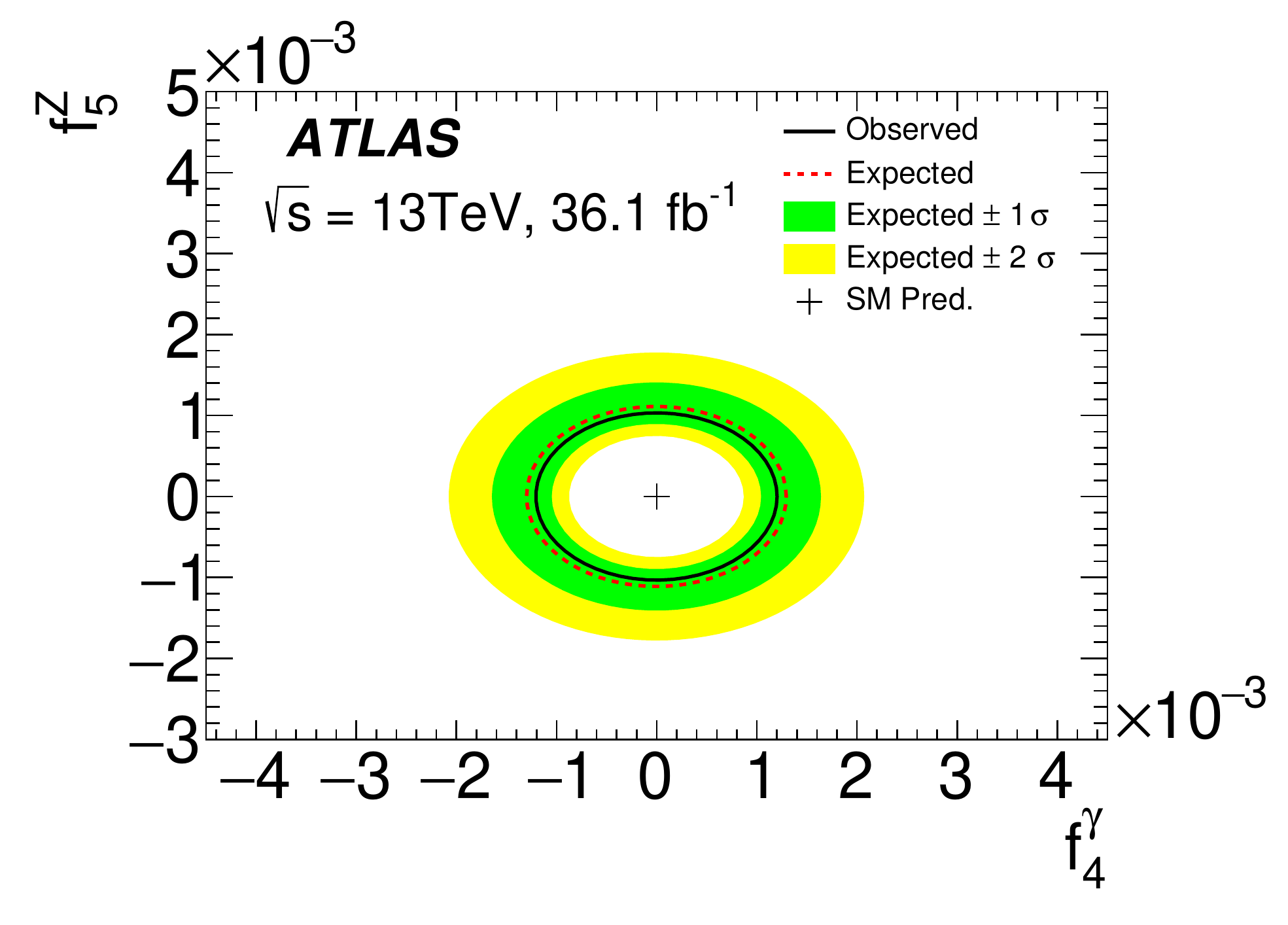}\label{fig:aTGC_f4g_f5Z_2D}
\includegraphics[width=0.495\textwidth]{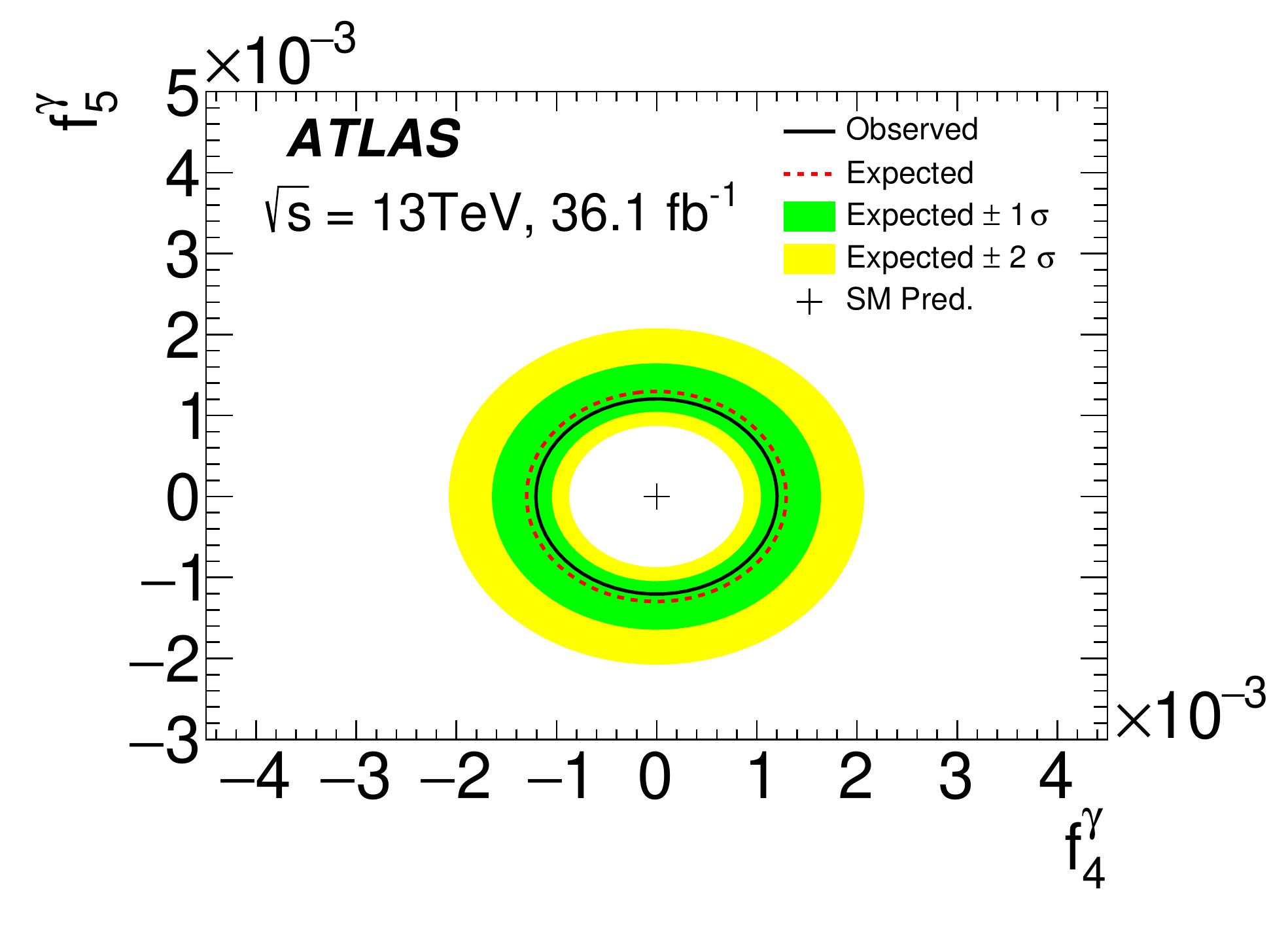}\label{fig:aTGC_f4g_f5g_2D}
\includegraphics[width=0.495\textwidth]{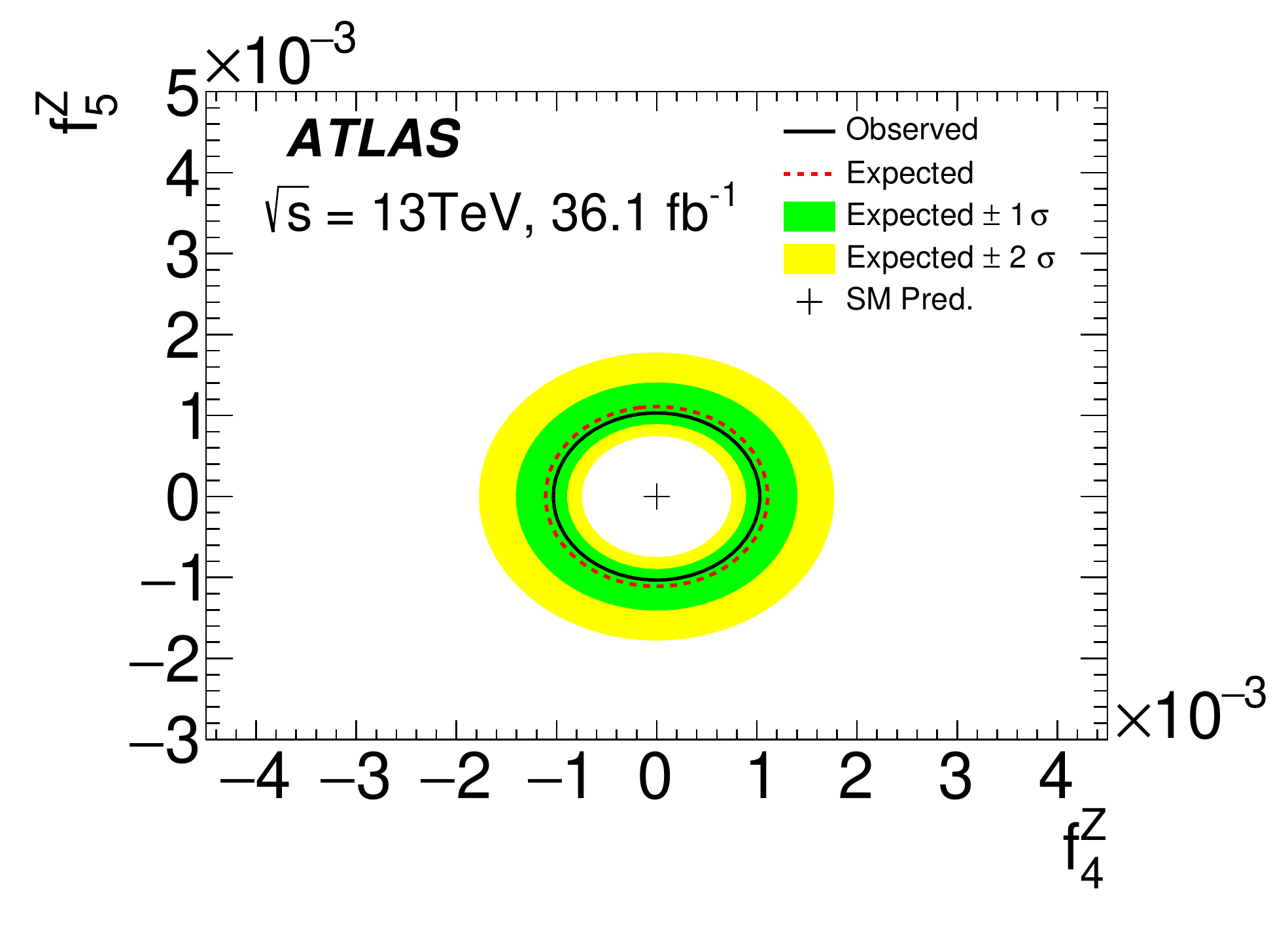}\label{fig:aTGC_f4Z_f5Z_2D}
\includegraphics[width=0.495\textwidth]{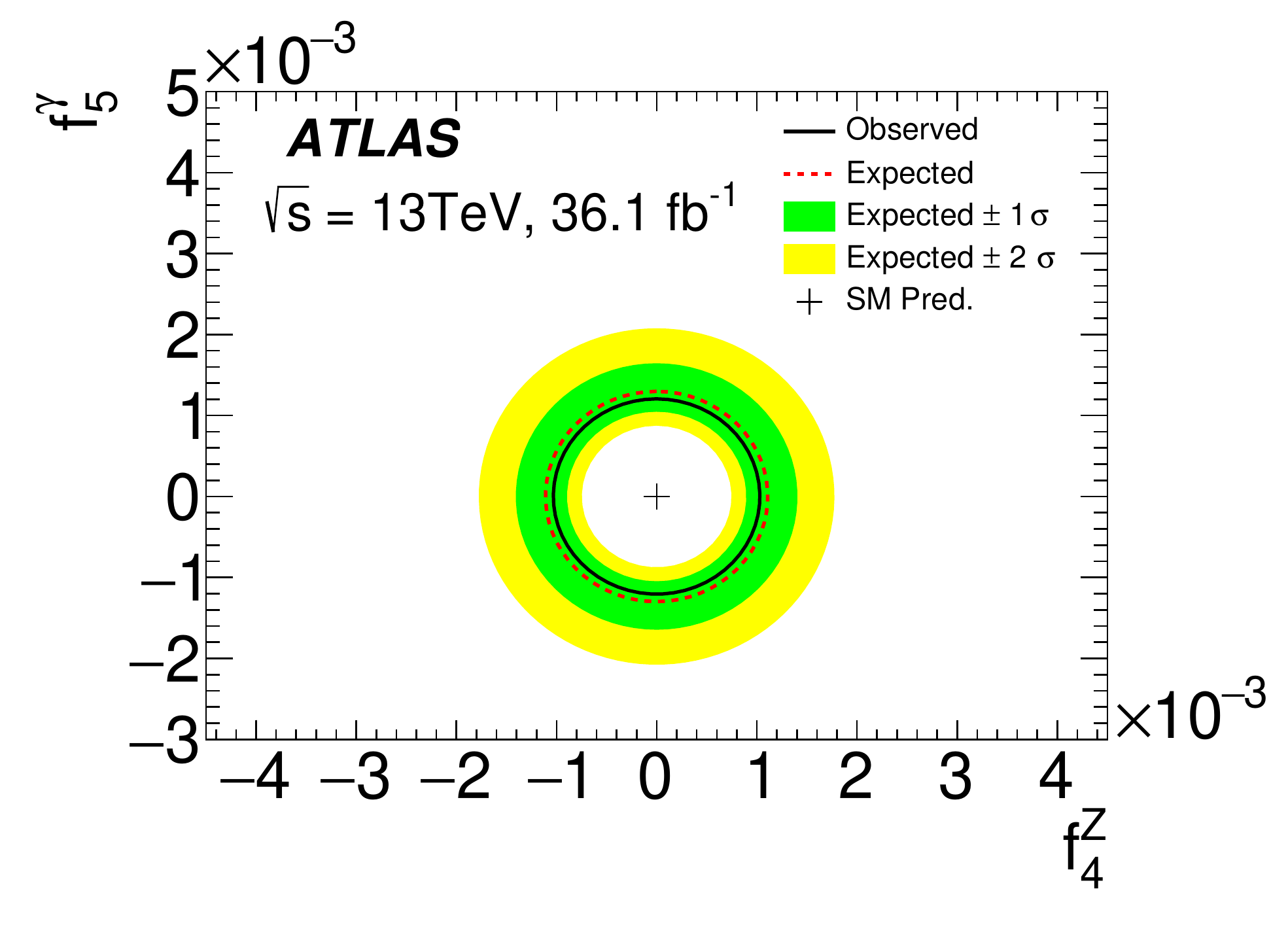}\label{fig:aTGC_f4Z_f5g_2D}
\includegraphics[width=0.495\textwidth]{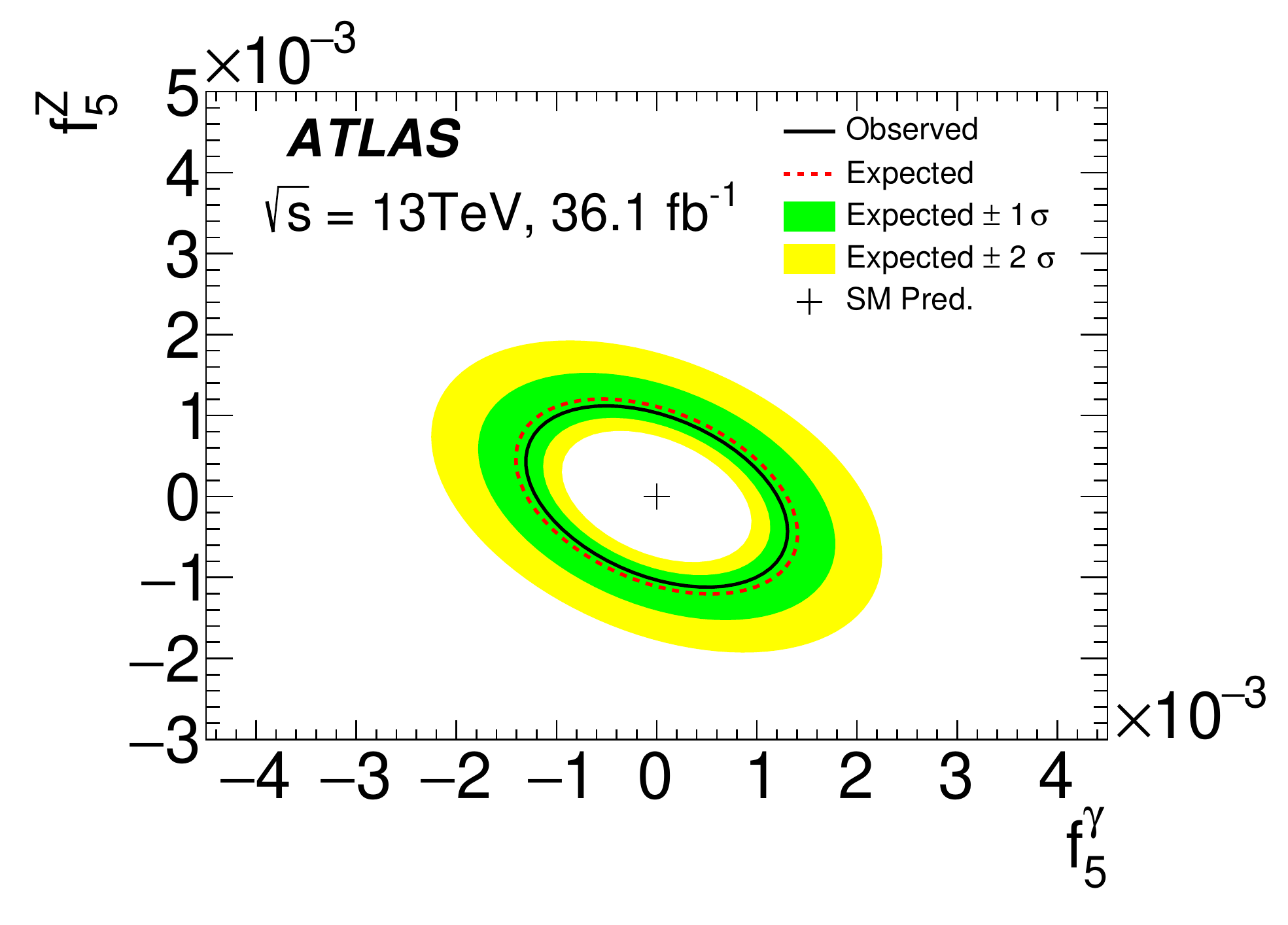}\label{fig:aTGC_f5g_f5Z_2D}
\end{center}
\caption{Expected and observed regions excluded at 95\% confidence level as a function of two aTGC parameters.
Confidence intervals for each pair of coupling parameters are derived with the other two parameters set to zero.
}
\label{fig:aTGC_2D_Limits}
\end{figure}
\FloatBarrier

 
\section{Conclusion}
\label{sec:conclusion}
 
This paper presents a measurement of $ZZ$ production using the \llnn{} final state, based on 36.1~\ifb{}
of data collected by the ATLAS detector at the LHC in $pp$ collisions at $\sqrt{s} =$ 13~\TeV.
The fiducial cross-section of the combined $ee$ and $\mu\mu$ channels is measured to be \xsfid{} =
25.4 $\pm$ 1.4 (stat) $\pm$ 0.9 (syst) $\pm$ 0.5 (lumi)~fb,
in agreement with the SM prediction of $22.4 \pm 1.3$~fb.
The integrated cross-sections in the total and
fiducial phase spaces are measured with a total uncertainty of 7\%,
which is significantly better than the previous measurement using the 8 \TeV{} data.
The measured cross-sections are slightly larger than the SM predictions,
but the difference is not significant considering the measurement and prediction uncertainties.
Differential cross-sections are reported for eight different kinematic variables in the fiducial phase space,
and no significant deviation from the expectations is found.
The measured \ptll{} spectrum in the fiducial phase space is used to set limits on aTGCs,
and the obtained 95\% confidence intervals for the aTGC parameters are more stringent
than those derived from the $4\ell$ channel.

\section*{Acknowledgements}
 
 
We thank CERN for the very successful operation of the LHC, as well as the
support staff from our institutions without whom ATLAS could not be
operated efficiently.
 
We acknowledge the support of ANPCyT, Argentina; YerPhI, Armenia; ARC, Australia; BMWFW and FWF, Austria; ANAS, Azerbaijan; SSTC, Belarus; CNPq and FAPESP, Brazil; NSERC, NRC and CFI, Canada; CERN; CONICYT, Chile; CAS, MOST and NSFC, China; COLCIENCIAS, Colombia; MSMT CR, MPO CR and VSC CR, Czech Republic; DNRF and DNSRC, Denmark; IN2P3-CNRS, CEA-DRF/IRFU, France; SRNSFG, Georgia; BMBF, HGF, and MPG, Germany; GSRT, Greece; RGC, Hong Kong SAR, China; ISF and Benoziyo Center, Israel; INFN, Italy; MEXT and JSPS, Japan; CNRST, Morocco; NWO, Netherlands; RCN, Norway; MNiSW and NCN, Poland; FCT, Portugal; MNE/IFA, Romania; MES of Russia and NRC KI, Russian Federation; JINR; MESTD, Serbia; MSSR, Slovakia; ARRS and MIZ\v{S}, Slovenia; DST/NRF, South Africa; MINECO, Spain; SRC and Wallenberg Foundation, Sweden; SERI, SNSF and Cantons of Bern and Geneva, Switzerland; MOST, Taiwan; TAEK, Turkey; STFC, United Kingdom; DOE and NSF, United States of America. In addition, individual groups and members have received support from BCKDF, CANARIE, CRC and Compute Canada, Canada; COST, ERC, ERDF, Horizon 2020, and Marie Sk{\l}odowska-Curie Actions, European Union; Investissements d' Avenir Labex and Idex, ANR, France; DFG and AvH Foundation, Germany; Herakleitos, Thales and Aristeia programmes co-financed by EU-ESF and the Greek NSRF, Greece; BSF-NSF and GIF, Israel; CERCA Programme Generalitat de Catalunya, Spain; The Royal Society and Leverhulme Trust, United Kingdom.
 
The crucial computing support from all WLCG partners is acknowledged gratefully, in particular from CERN, the ATLAS Tier-1 facilities at TRIUMF (Canada), NDGF (Denmark, Norway, Sweden), CC-IN2P3 (France), KIT/GridKA (Germany), INFN-CNAF (Italy), NL-T1 (Netherlands), PIC (Spain), ASGC (Taiwan), RAL (UK) and BNL (USA), the Tier-2 facilities worldwide and large non-WLCG resource providers. Major contributors of computing resources are listed in Ref.~\cite{ATL-GEN-PUB-2016-002}.
 

\printbibliography
 
\clearpage 
 
\begin{flushleft}
{\Large The ATLAS Collaboration}

\bigskip

M.~Aaboud$^\textrm{\scriptsize 35d}$,    
G.~Aad$^\textrm{\scriptsize 101}$,    
B.~Abbott$^\textrm{\scriptsize 128}$,    
D.C.~Abbott$^\textrm{\scriptsize 102}$,    
O.~Abdinov$^\textrm{\scriptsize 13,*}$,    
A.~Abed~Abud$^\textrm{\scriptsize 70a,70b}$,    
D.K.~Abhayasinghe$^\textrm{\scriptsize 93}$,    
S.H.~Abidi$^\textrm{\scriptsize 167}$,    
O.S.~AbouZeid$^\textrm{\scriptsize 40}$,    
N.L.~Abraham$^\textrm{\scriptsize 156}$
% ...
\bigskip
\\

% ...
$^{13}$Institute of Physics, Azerbaijan Academy of Sciences, Baku; Azerbaijan.\\
% ...
$^{35}$$^{(a)}$Facult\'e des Sciences Ain Chock, R\'eseau Universitaire de Physique des Hautes Energies - Universit\'e Hassan II, Casablanca;$^{(b)}$Facult\'{e} des Sciences, Universit\'{e} Ibn-Tofail, K\'{e}nitra;$^{(c)}$Facult\'e des Sciences Semlalia, Universit\'e Cadi Ayyad, LPHEA-Marrakech;$^{(d)}$Facult\'e des Sciences, Universit\'e Mohamed Premier and LPTPM, Oujda;$^{(e)}$Facult\'e des sciences, Universit\'e Mohammed V, Rabat; Morocco.\\
% ...
$^{40}$Niels Bohr Institute, University of Copenhagen, Copenhagen; Denmark.\\
% ...
$^{70}$$^{(a)}$INFN Sezione di Pavia;$^{(b)}$Dipartimento di Fisica, Universit\`a di Pavia, Pavia; Italy.\\
% ...
$^{93}$Department of Physics, Royal Holloway University of London, Egham; United Kingdom.\\
% ...
$^{101}$CPPM, Aix-Marseille Universit\'e, CNRS/IN2P3, Marseille; France.\\
$^{102}$Department of Physics, University of Massachusetts, Amherst MA; United States of America.\\
% ...
$^{128}$Homer L. Dodge Department of Physics and Astronomy, University of Oklahoma, Norman OK; United States of America.\\
% ...
$^{156}$Department of Physics and Astronomy, University of Sussex, Brighton; United Kingdom.\\
% ...
$^{167}$Department of Physics, University of Toronto, Toronto ON; Canada.\\
% ...
$^{a}$ Also at Centre for High Performance Computing, CSIR Campus, Rosebank, Cape Town; South Africa.\\
$^{b}$ Also at CERN, Geneva; Switzerland.\\
$^{c}$ Also at CPPM, Aix-Marseille Universit\'e, CNRS/IN2P3, Marseille; France.\\
$^{d}$ Also at D\'epartement de Physique Nucl\'eaire et Corpusculaire, Universit\'e de Gen\`eve, Gen\`eve; Switzerland.\\
$^{e}$ Also at Departament de Fisica de la Universitat Autonoma de Barcelona, Barcelona; Spain.\\
% ...
$^{*}$ Deceased

\end{flushleft}


-------------------------------------------------------------------------------

\end{document}